%% file: main.tex
\documentclass[hidelinks,12pt]{article}
\usepackage[export]{adjustbox} 
\usepackage{scicite}
\usepackage{color}
\usepackage[dvipsnames]{xcolor}
\usepackage{graphicx}	
\usepackage{amssymb}
\usepackage{enumitem}
\usepackage{bm}
\usepackage{caption}
\usepackage{subcaption}
\usepackage{mathtools}
\usepackage{titlesec}
\usepackage{bbm}

\usepackage{colortbl}
\definecolor{Gray}{gray}{0.9}
\usepackage{hyperref}
\makeatletter
\renewcommand*{\@biblabel}[1]{\hfill#1.}
\makeatother
\newcounter{myequation}
\makeatletter
\@addtoreset{equation}{myequation}
\makeatother
\makeatletter
\renewcommand*{\@textcolor}[3]{%
\protect\leavevmode
\begingroup
\color#1{#2}#3%
\endgroup
}
\makeatother

\usepackage{times}
\emergencystretch=\maxdimen
\usepackage[labelsep=period]{caption}
\usepackage[labelfont=bf]{caption}
\usepackage{amsmath,accents}
\usepackage[para,online,flushleft]{threeparttable}
\usepackage{multirow}

\newenvironment{sciabstract}{%
\begin{quote} \bf}
{\end{quote}}

\renewcommand{\figurename}{Fig.}

\title{\textbf{Data-Driven, Physics-Informed Descriptors of Cation Ordering in Multicomponent Oxides}}

\author{
\hspace{0mm}Jiayu Peng,$^{1}$ James Damewood,$^{1,2}$ Rafael Gómez-Bombarelli$^{1,*}$\\
\\
\hspace{0mm}\normalsize{$^{1}$Department of Materials Science and Engineering, Massachusetts Institute of Technology,}\\
\hspace{0mm}\normalsize{Cambridge, MA 02139, USA}\\
\hspace{0mm}\normalsize{$^{2}$Center for Computational Science and Engineering, Massachusetts Institute of Technology,}\\
\hspace{0mm}\normalsize{Cambridge, MA 02139, USA}\\
\hspace{0mm}{$^{*}$\small Correspondence: rafagb@mit.edu}\\
}

\date{}

\begin{document}


\maketitle 

\baselineskip18pt

\vspace{-2mm}\noindent{\bf Summary:}
Data-driven, physically interpretable descriptors of cation ordering are established for multicomponent perovskite oxides.

\phantomsection
\addcontentsline{toc}{section}{Abstract}
\begin{sciabstract}

The structural tunability and compositional diversity of multicomponent perovskite oxides have enabled their various applications, including catalysis and electronics. The cation ordering in these oxides, ranging from disordered (i.e., high-entropy) to ordered (e.g., rocksalt), profoundly influences their properties. While computational design tools can typically predict properties associated with a particular ordering, inferring which ordering---if any---will be observed in synthesized oxides remains challenging. Here, we leveraged first-principles simulations and machine learning to develop data-driven, physics-informed descriptors of experimental ordering in multicomponent perovskites and compared them with traditional physicochemical descriptors, e.g., ionic radii and oxidation states. The fitted low-dimensional classification models correctly rank up to $93\%$ of compositions in an experimental dataset of $190$ perovskites between cation-ordered and disordered, offering a rigorous benchmark between theory and experiments. Furthermore, these descriptors accelerate high-throughput virtual screening of multicomponent oxides by predicting their dominant ordering to avoid costly, exhaustive simulations of cation arrangements.

\end{sciabstract}

\clearpage

\phantomsection
\addcontentsline{toc}{section}{Introduction}
\section*{Introduction}

Multicomponent perovskite oxides are an essential class of inorganic materials with diverse key applications, ranging from electronics to renewable energy \cite{Hwang:2017,Pena:2002}. Having a general formula of ABO\textsubscript{3}, these oxides are composed of a network of B-site cations and oxygen anions, forming corner-sharing BO\textsubscript{6} octahedra surrounded by A-site cations. Multiple B-site and A-site cations can exist in such oxides, and these cations almost span the entire periodic table. The compositional flexibility of multicomponent perovskite oxides has enabled their various functions, including high-temperature superconductivity for electronics \cite{MacManus-Driscoll:2021}, colossal magnetoresistance for magnetic devices\cite{Salamon:2001}, Li-ion conductivity for batteries \cite{Bachman:2016}, and surface reactivity for heterogeneous catalysis \cite{Hwang:2021} and electrocatalysis \cite{Kuznetsov:2020,Peng:2022}.

Concomitant with their compositional versatility, multicomponent perovskite oxides exhibit variability in the distribution and ordering of different cations. The specific arrangements of cation sublattices that are periodically repeated in these oxides can have a profound influence on their physical and chemical properties. For example, Cu cation layers in cuprate perovskites are crucial for enabling high-temperature superconductivity \cite{Dagotto:1994}. Moreover, cation order-to-disorder tuning in perovskite oxides has been widely leveraged to optimize their optical absorption \cite{Nechache:2015}, magnetic behaviors \cite{Senn:2014}, ion mobility \cite{Stramare:2003,Uberuaga:2015}, catalytic activity \cite{Grimaud:2013}, and electromechanical properties \cite{Grinberg:2002}, facilitating their atom-by-atom design for diverse practical applications.

Unfortunately, the design of multicomponent perovskite oxides has been severely impeded by a lack of physical principles to universally rationalize and accurately predict their cation ordering. Although a few empirical rules have been established to correlate the experimental cation ordering in these oxides with the atomic parameters of their constituent ions \cite{Anderson:1993,King:2010}, such rules have often been shown to fail across broad oxide space for correctly classifying perovskite oxides into cation-ordered and disordered structures \cite{Vasala:2015}. For instance, increasing the charge and size differences between B-site cations in multicomponent perovskite oxides has been reported to favor B-site sublattices with rocksalt ordering over their cation-disordered counterparts \cite{Anderson:1993}. However, while these correlations can be explained by a simple hypothesis that the rocksalt ordering is energetically favored to electrostatically maximize the separation of the more highly charged B-site cations \cite{King:2010}, it cannot quantitatively distinguish cation-ordered perovskite oxides from their disordered counterparts \cite{Vasala:2015}. Similarly, having a large charge mismatch between A-site ions has been proposed to give rise to A-site layered ordering \cite{King:2010}, but this trend has limited predictive power \cite{Ghosh:2022}. These observations corroborate recent discoveries that many traditional empirical principles predicting phase stability and competition in oxides \cite{Goldschmidt:1926,Pauling:1929} no longer hold true when re-examined across a vast compositional space \cite{George:2020,Bartel:2019}. It is thus critical to rationalize the limitation of such conventional empirical rules and develop new physics-driven design principles that exhibit greater generalizability and higher accuracy for inferring the cation ordering in multicomponent perovskite oxides.

High-throughput atomistic simulations, possibly accelerated with machine learning, could provide highly accurate, physics-informed design principles of cation ordering across a wide compositional space of multicomponent perovskites based on formation energetics \cite{Ghosh:2022,Raabe:2023,Ferrari:2023}. Nevertheless, this exploration of ordering is constrained by a lack of systematic benchmarks between theory and experiments. Although density functional theory (DFT) calculations have been shown to broadly and accurately reproduce the compositional dependence of material properties \cite{Jain:2013,Hautier:2012,Peng:2022}, previous computational studies on cation ordering have merely evaluated a handful of oxides by comparing DFT-computed energetics with their experimental cation arrangements \cite{Grinberg:2002,Burton:1999,Gou:2017,Paul:2020,Shaikh:2021,Kaczkowski:2022,Bhuyan:2022}. Moreover, DFT-backed, data-driven models have been designed to infer the formation energetics of oxide structures with different cation arrangements. However, these models suffer from drawbacks: they are not physically interpretable since they just map structures to energies; they are trained with DFT results, lacking comprehensive experimental validation; and they require relatively large training datasets, barring the incorporation of the much more limited experimental data \cite{Peng2:2022,Ye:2018,Barroso-Luque:2022,Yuan:2023}. Thus, in order to establish descriptors of cation ordering, it is crucial to systematically examine whether DFT can capture experimental cation ordering in a large compositional space of multicomponent oxides and compare or combine these DFT energetics with descriptor-powered, interpretable machine learning models. 

The dearth of cation ordering descriptors has further hindered the high-throughput virtual screening (HTVS) of multicomponent perovskite oxides. In HTVS, it has been particularly challenging to representatively, yet efficiently navigate the immense configurational space of cation ordering. Currently, the HTVS of such complex oxides is computationally inefficient, requiring the costly DFT sampling of large supercells \cite{Ferrari:2023,Jiang:2016} with randomly selected \cite{Choubisa:2023,Tran:2023} or manually predefined cation arrangements \cite{Ma:2021,Jacobs:2022}. These existing sampling strategies are not exhaustive, so they lower the otherwise prohibitive cost of extensively calculating all possible configurations with DFT. However, it is unclear whether such strategies systematically capture the low-energy cation arrangements of multicomponent oxides or, more importantly, whether they reflect experimentally observed ordering. Therefore, it is imperative to use cation ordering descriptors to boost HTVS by optimizing the balance between the breadth and cost of DFT sampling in exploring the vast configurational space of cation arrangements. 

Here, combining high-throughput DFT calculations and machine learning, we developed data-driven, physics-informed descriptors of cation ordering to universally and accurately infer experimental cation ordering in multicomponent perovskite oxides. By constructing a dataset of $190$ experimental double perovskites with symmetrically inequivalent cation arrangements, we showed that these new descriptors significantly outperform their traditional counterparts, e.g., the charge and size differences of B-site ions \cite{Anderson:1993,King:2010,Vasala:2015}, correctly ranking up to $93\%$ oxide compositions between cation-ordered and disordered. Such descriptors can be rationalized as physically interpretable parameters, including the thermodynamic likelihood of each ordering and configurational entropy, which can be obtained from the non-exhaustive DFT calculations of only a few simple model cation arrangements. These results use abundant prior experiments to offer a rigorous validation of our simulation and data science strategies. By coupling these descriptors with HTVS, we demonstrated that descriptor-driven DFT sampling strategies reside at the Pareto front of accuracy--cost trade-offs when navigating the configurational space of cation arrangements, with comparable computational cost but lower sampling error than their descriptor-free counterparts. This work highlights an effective paradigm for understanding and predicting chemical and structural orders in multicomponent materials to accelerate their data-driven discoveries across a massive compositional space.

\phantomsection
\addcontentsline{toc}{section}{Results}
\section*{Results}

\phantomsection
\addcontentsline{toc}{subsection}{Building physics-informed descriptors to infer cation ordering}
\subsection*{Building physics-informed descriptors to infer cation ordering}

\begin{figure}[hbt!]
\phantomsection
\begin{center}
\includegraphics[max size={\textwidth}{\textheight}]{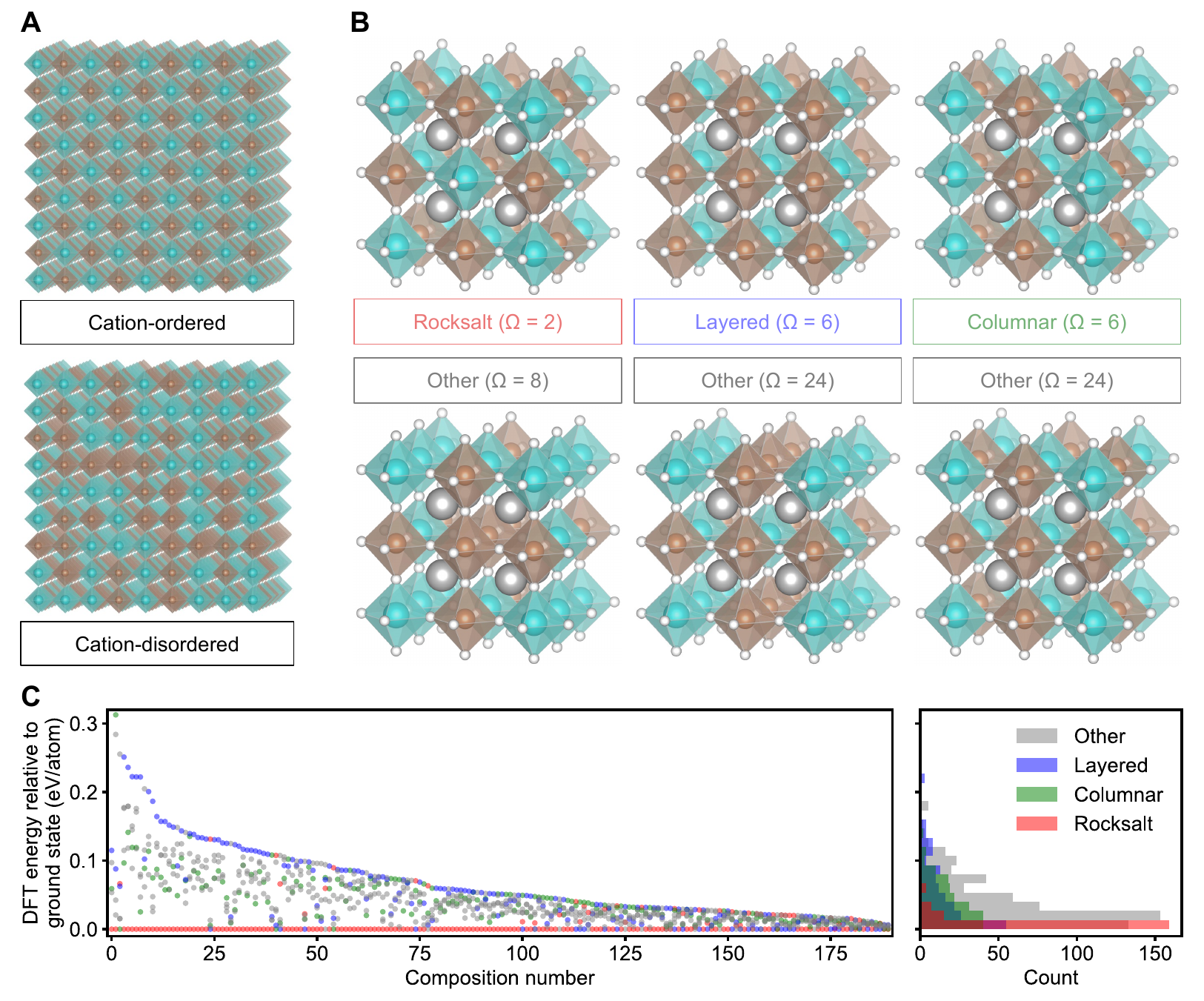}
\caption{
\textbf{Formation energetics of model cation arrangements.}
(\textbf{A}) Schematic structures of cation-ordered and disordered multicomponent perovskite oxides, represented by $8 \times 8 \times 8$ supercells. (\textbf{B}) Model structures of six symmetrically inequivalent cation arrangements in a $2 \times 2 \times 2$ supercell of A\textsubscript{2}B'B"O\textsubscript{6} perovskites. $\Omega$ is the degeneracy of cation arrangements. Color scheme: A, grey; B', brown; B", cyan; O, white. (\textbf{C}) DFT-computed formation energies of the six model structures relative to that of the computationally ground-state cation arrangement for an experimental dataset of $190$ A\textsubscript{2}B'B"O\textsubscript{6} perovskites.
}
\label{fig:Main_dataset_statistics}
\end{center}
\end{figure}

To enable a systematic benchmark between theory and experiments and develop physics-driven descriptors of cation ordering, we constructed a compositionally diverse DFT dataset of $190$ experimental multicomponent perovskite oxides (Figs. \ref{fig:Main_dataset_statistics} and \ref{fig:SI_elemental_info}; Materials and Methods). In this dataset, we focused on A\textsubscript{2}B'B"O\textsubscript{6} double perovskites to establish and evaluate descriptors, as A\textsubscript{2}B'B"O\textsubscript{6} compositions have dominated the literature with experimentally quantified cation ordering \cite{Vasala:2015}. Moreover, we examined selected multicomponent perovskites with less common compositions and orderings \cite{Azuma:1998,Aimi:2014,Yamada:2008,Long:2009,Byeon:2005} in detail (Table \ref{table:hold_out_test_set}) as an additional evaluation set to further assess the accuracy and generalizability of these descriptors.

With the datasets, we proposed physically interpretable descriptors of cation ordering based on the formation energetics of simple model cation arrangements. While large supercells have been widely utilized to capture the quasirandom cation sublattices of multicomponent oxides (Fig. \ref{fig:Main_dataset_statistics}A), these massive supercells can be too computationally expensive for high-throughput investigation \cite{Zhang:2023}. To tackle this limitation, we examined the formation energetics of simple, less expensive $2 \times 2 \times 2$ supercells with symmetrically distinct cation arrangements, offering a computationally efficient approach for navigating the energy landscape of cation ordering in complex perovskites. Specifically, distributing four B' and four B" cations among the eight B sites in $2 \times 2 \times 2$ A\textsubscript{2}B'B"O\textsubscript{6} supercells leads to as many as $(_4^8) = 70$ combinations, but such combinations can be reduced to only six symmetrically inequivalent configurations (Fig. \ref{fig:Main_dataset_statistics}B). Among the six configurations, three of them have been named \textit{layered}, \textit{columnar}, and \textit{rocksalt} arrangements, with cations alternating in one, two, and three dimensions, respectively \cite{Anderson:1993,King:2010,Vasala:2015}. Computationally, the rocksalt arrangement was found to be the ground-state configuration for most of these $190$ oxides (Fig. \ref{fig:Main_dataset_statistics}C), consistent with prior experimental observations that, under ambient pressure, all but one known cation-ordered A\textsubscript{2}B'B"O\textsubscript{6} oxides have B-site sublattices with rocksalt ordering \cite{Anderson:1993,Vasala:2015}. Although such a rocksalt arrangement tends to be the ground-state configuration, other cation arrangements can also be thermodynamically favorable if their formation energies are only slightly higher than their ground-state counterpart, especially if they have favorable configurational entropy. Thus, the tendency to deviate from randomness may be characterized by the energy gap between ground-state and higher-energy cation arrangements ($\Delta E$) or, more rigorously, the thermodynamic likelihood of forming a configuration $i$:
\begin{equation}
\label{eqn:thermo_prob}
P_i = \Omega_i \frac{\exp{\left( -\frac{E_i}{k_{\mathrm{B}}T} \right)}}{\sum_{j=1}^{N_{\mathrm{unique}}} \Omega_j \exp{\left( -\frac{E_j}{k_{\mathrm{B}}T} \right)}}, \quad i=1,\ldots,N_{\mathrm{unique}}
\end{equation}
where $E_i$ is the formation energy of the cation configuration $i$ (Fig. \ref{fig:Main_dataset_statistics}C), $\Omega_i$ is the degeneracy of this configuration (Fig. \ref{fig:Main_dataset_statistics}B), $N_{\mathrm{unique}}$ is the number of symmetrically unique configurations ($N_{\mathrm{unique}} = 6$ for A\textsubscript{2}B'B"O\textsubscript{6}), $k_{\mathrm{B}}$ is the Boltzmann constant, and $T$ is the temperature. Another parameter that potentially captures the general degree of order or disorder is the configurational entropy, which weights across all possible orderings:
\begin{equation}
\label{eqn:conf_entropy}
S_{\mathrm{conf}} = -\sum_{i=1}^n \frac{\Omega_i}{\ln{\sum_{k=1}^{N_{\mathrm{unique}}} \Omega_k}} \frac{\exp{\left( -\frac{E_i}{k_{\mathrm{B}}T} \right)}}{\sum_{j=1}^{N_{\mathrm{unique}}} \Omega_j \exp{\left( -\frac{E_j}{k_{\mathrm{B}}T} \right)}} \ln{ \frac{\exp{\left( -\frac{E_i}{k_{\mathrm{B}}T} \right)}}{\sum_{j=1}^{N_{\mathrm{unique}}} \Omega_j \exp{\left( -\frac{E_j}{k_{\mathrm{B}}T} \right)}} }
\end{equation}
which has a range between $S_{\mathrm{residual}}$ and $1$, where $S_{\mathrm{residual}}$ is the residual entropy stemming from the configurational multiplicity of the symmetrically unique ground-state arrangement (Eq. \ref{eqn:descriptor_construction_residual_entropy}; Materials and Methods). Intuitively, the difference between $S_{\mathrm{residual}}$ and $S_{\mathrm{conf}}$ quantifies whether low-lying alternative orderings to the ground state are expected to be populated, beyond the sheer configurational entropy of the degenerate arrangements of the ground state ($S_{\mathrm{residual}}$). Overall, these three parameters, $\Delta E$, $P_i$, and $S_{\mathrm{conf}}$, characterize the statistical thermodynamics of various cation configurations in multicomponent oxides and can thus serve as physics-driven descriptors to infer their experimental cation ordering.

\phantomsection
\addcontentsline{toc}{subsection}{Evaluating cation ordering descriptors with experimental datasets}
\subsection*{Evaluating cation ordering descriptors with experimental datasets}

\begin{figure}[hbt!]
\phantomsection
\begin{center}
\includegraphics[max size={\textwidth}{\textheight}]{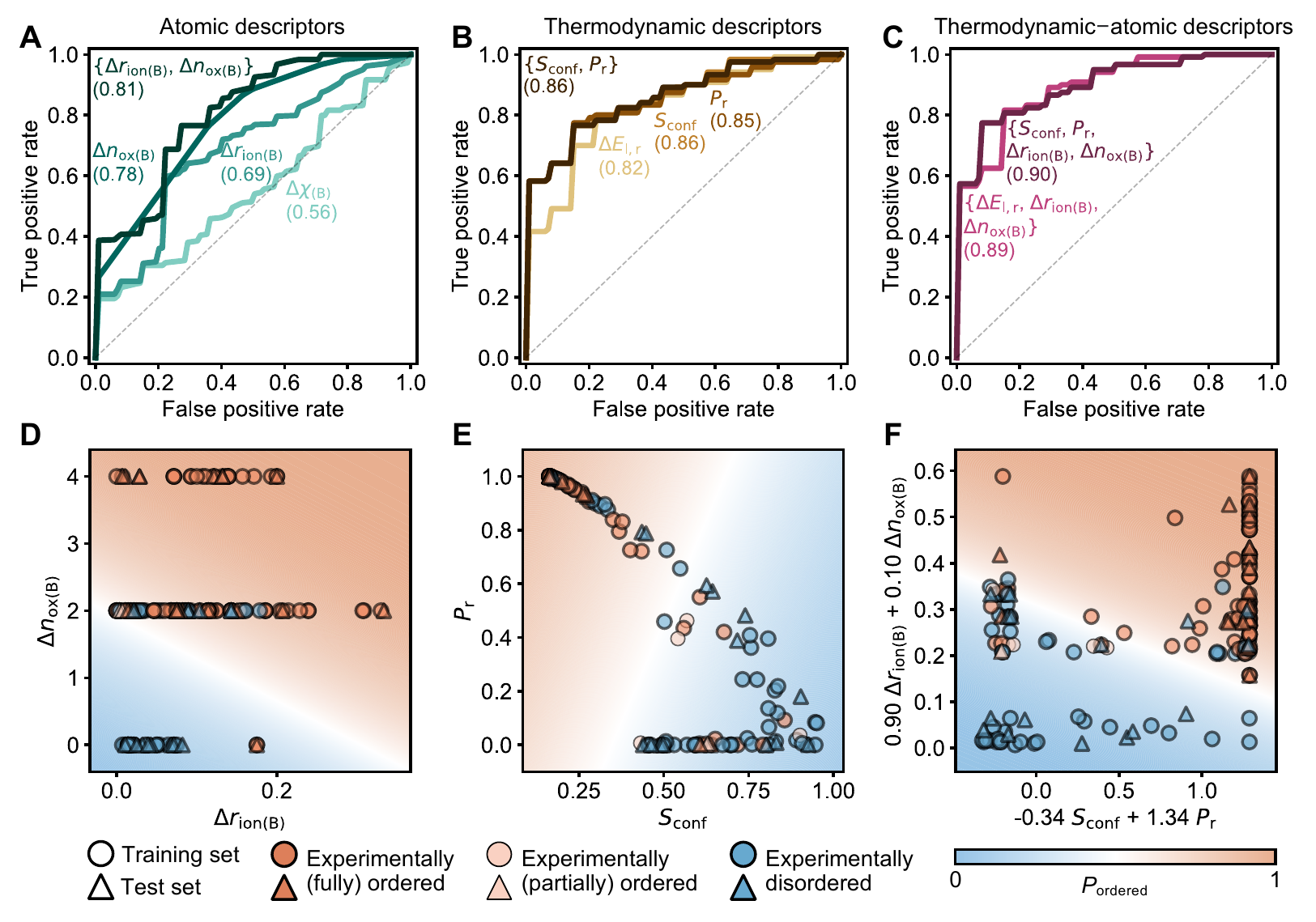}
\caption{
\textbf{Descriptor performance for classifying A\textsubscript{2}B'B"O\textsubscript{6} perovskites as cation-ordered or disordered.}
(\textbf{A} to \textbf{C}) Performance of one- and multi-dimensional descriptors using logistic regression with five-fold cross-validation, quantified by ROC curves with AUC in parentheses. Dashed lines show the ROC curves of a random classifier (AUC $= 0.5$). $\Delta \chi_{\mathrm{(B)}}$, $\Delta r_{\mathrm{ion(B)}}$, and $\Delta n_{\mathrm{ox(B)}}$ are the differences between the electronegativities, ionic radii, and oxidation states of B-site ions, respectively. $\Delta E_{\mathrm{l,r}}$, $S_{\mathrm{conf}}$, and $P_{\mathrm{r}}$ are the energy gap between the layer and rocksalt arrangements, the configuration entropy (Eq. \ref{eqn:conf_entropy}), and the likelihood of generating the rocksalt configuration (Eq. \ref{eqn:thermo_prob}), respectively. (\textbf{D} to \textbf{F}) Decision boundaries of top-performing multi-dimensional descriptors. $P_{\mathrm{ordered}}$ is the predicted likelihood that an oxide should be classified as cation-ordered through logistic regression.
}
\label{fig:Main_binary_classfication}
\end{center}
\end{figure}

We compared these physically interpretable thermodynamic parameters with traditional atomic descriptors \cite{Anderson:1993,King:2010,Vasala:2015} to investigate the performance in classifying multicomponent perovskite oxides as cation-ordered or disordered across a wide oxide space. Specifically, for the dataset of $190$ A\textsubscript{2}B'B"O\textsubscript{6} oxides (Fig. \ref{fig:Main_dataset_statistics}), we established one- or multi-dimensional descriptors based on the combination of these parameters and fitted a logistic classifier to predict whether such oxides are experimentally cation-disordered or ordered with B-site rocksalt ordering (Fig. \ref{fig:Main_binary_classfication}; Materials and Methods). The descriptor performance was assessed by the confusion matrices and the receiver operating characteristic (ROC) curves. While a confusion matrix characterizes the classification accuracy at the optimized decision boundary, the area under the curve (AUC) in ROC curves quantifies the overall accuracy at various classification thresholds, providing a comprehensive, systematic evaluation of the descriptor performance, and is a more preferred metric for imbalanced classes. Intuitively, the ROC--AUC of a descriptor can be understood as the average accuracy (or probability) that this descriptor ranks an experimentally cation-ordered perovskite oxide as higher-likelihood of being ordered than a cation-disordered one.

Increasing $S_{\mathrm{conf}}$ and lowering the likelihood of forming the rocksalt cation configuration ($P_{\mathrm{r}}$) were found to markedly outperform traditional atomic descriptors in favoring disordered B-site sublattices over cation-ordered ones. Specifically, while decreasing the charge ($\Delta n_{\mathrm{ox(B)}}$) and size differences between B-site ions ($\Delta r_{\mathrm{ion(B)}}$) has been widely reported to lead to B-site sublattices with cation disorder \cite{Anderson:1993,King:2010,Vasala:2015}, these two parameters can hardly effectively classify experimentally cation-ordered and disordered oxides. $\Delta n_{\mathrm{ox(B)}}$ and $\Delta r_{\mathrm{ion(B)}}$ have ROC--AUC scores of $0.78$ and $0.69$, respectively, and coupling them as a two-dimensional (2D) descriptor results in an AUC of $0.81$ (Fig. \ref{fig:Main_binary_classfication}A), indicating an $81\%$ accuracy in ranking cation-ordered oxides above their disordered counterparts. Nevertheless, although this 2D descriptor behaves better than other atomic parameters such as the electronegativity difference ($\Delta \chi_{\mathrm{(B)}}$), which has a ROC--AUC comparable to a trivial random classifier (Fig. \ref{fig:Main_binary_classfication}A), it cannot quantitatively differ cation-disordered perovskites from ordered oxides in its 2D parameter space (Fig. \ref{fig:Main_binary_classfication}D). The limitation of $\Delta n_{\mathrm{ox(B)}}$ and $\Delta r_{\mathrm{ion(B)}}$ is corroborated by their confusion matrices, where more than half of experimentally cation-disordered perovskites are misclassified as ordered (Fig. \ref{fig:SI_additional_descriptor_performance}). In contrast, thermodynamic parameters, including $P_{\mathrm{r}}$ (Eq. \ref{eqn:thermo_prob}) and $S_{\mathrm{conf}}$ (Eq. \ref{eqn:conf_entropy}), lead to a 2D descriptor that has an AUC of $0.86$ (Fig. \ref{fig:Main_binary_classfication}B) and can effectively distinguish cation-ordered oxides from their disordered counterparts (Fig. \ref{fig:Main_binary_classfication}E), highlighting the power of thermodynamic descriptors in understanding and predicting experimental cation ordering. 

Hybridizing thermodynamic and atomic parameters further contributes to highly predictive descriptors of cation ordering. Particularly, $\{S_{\mathrm{conf}}, P_{\mathrm{r}}, \Delta r_{\mathrm{ion(B)}}, \Delta n_{\mathrm{ox(B)}}\}$ serves as an effective four-dimensional (4D) descriptor with a ROC--AUC of $0.90$ (Fig. \ref{fig:Main_binary_classfication}C). Moreover, replacing $S_{\mathrm{conf}}$ and $P_{\mathrm{r}}$ with the energy gap between the layer and rocksalt configurations ($\Delta E_{\mathrm{l,r}}$, Fig. \ref{fig:Main_dataset_statistics}) generates a comparably accurate three-dimensional (3D) descriptor with an AUC of $0.89$. The comparable performance of such hybrid descriptors can be explained by the strong influence of $\Delta E_{\mathrm{l,r}}$ on $S_{\mathrm{conf}}$ and $P_{\mathrm{r}}$. Moreover, the higher accuracy of $\{S_{\mathrm{conf}}, P_{\mathrm{r}}, \Delta r_{\mathrm{ion(B)}}, \Delta n_{\mathrm{ox(B)}}\}$ and $\{\Delta E_{\mathrm{l,r}}, \Delta r_{\mathrm{ion(B)}}, \Delta n_{\mathrm{ox(B)}}\}$ than thermodynamic or atomic descriptors separately can be justified by the orthogonality among these descriptors (Fig. \ref{fig:SI_correlation_matrix}). Intriguingly, when training the logistic regression model with $\{S_{\mathrm{conf}}, P_{\mathrm{r}}, \Delta r_{\mathrm{ion(B)}}, \Delta n_{\mathrm{ox(B)}}\}$ or $\{\Delta E_{\mathrm{l,r}}, \Delta r_{\mathrm{ion(B)}}, \Delta n_{\mathrm{ox(B)}}\}$, we did not provide the model with the information that eight out of the $190$ A\textsubscript{2}B'B"O\textsubscript{6} oxides are partially cation-ordered, but such descriptors appear to capture the experimentally partial cation ordering in these oxides by having their data close to the decision boundaries (Figs. \ref{fig:Main_binary_classfication}F and \ref{fig:SI_additional_descriptor_performance}K). Given such an experimental ambiguity, filtering out these eight experimentally partially cation-ordered perovskites can result in ROC--AUC scores of $0.93$ and $0.92$ for $\{S_{\mathrm{conf}}, P_{\mathrm{r}}, \Delta r_{\mathrm{ion(B)}}, \Delta n_{\mathrm{ox(B)}}\}$ and $\{\Delta E_{\mathrm{l,r}}, \Delta r_{\mathrm{ion(B)}}, \Delta n_{\mathrm{ox(B)}}\}$ (Fig. \ref{fig:SI_filtering_partially_experimentally_ordered}), respectively, demonstrating that such 4D and 3D hybrid descriptors have up to $93\%$ and $92\%$ accuracies in ranking cation-ordered A\textsubscript{2}B'B"O\textsubscript{6} perovskites above their disordered counterparts for unambiguous experimental data.

\begin{table}[hbt!]
\phantomsection
\centering
\begin{threeparttable}
\caption{
\textbf{Descriptor performance for perovskite oxides with exotic cation orderings.}
}
\label{table:hold_out_test_set}
\begin{tabular}{llcccc}
\hline
Composition & Experimental ordering & $N_{\mathrm{total}},\ N_{\mathrm{unique}}$ & $P_{\mathrm{exp}}$ & $S_{\mathrm{conf}},\ S_{\mathrm{residual}}$ \\
\hline
Nd\textsubscript{2}CuSnO\textsubscript{6} & B-site layered \cite{Azuma:1998} & $70,\ 6$ & $0.98$ & $0.46,\ 0.42$ \\
Pr\textsubscript{2}CuSnO\textsubscript{6} & B-site layered \cite{Azuma:1998} & $70,\ 6$ & $1.00$ & $0.42,\ 0.42$ \\
Sm\textsubscript{2}CuSnO\textsubscript{6} & B-site layered \cite{Azuma:1998} & $70,\ 6$ & $0.98$ & $0.45,\ 0.42$ \\
CaMnTi\textsubscript{2}O\textsubscript{6} & A-site columnar \cite{Aimi:2014} & $70,\ 6$ & $1.00$ & $0.43,\ 0.42$ \\
CaCu\textsubscript{3}Fe\textsubscript{4}O\textsubscript{12} & A-site body-centered \cite{Yamada:2008} & $28,\ 3$ & $1.00$ & $0.42,\ 0.42$ \\
LaCu\textsubscript{3}Fe\textsubscript{4}O\textsubscript{12} & A-site body-centered \cite{Long:2009} & $28,\ 3$ & $1.00$ & $0.42,\ 0.42$ \\
\multirow{2}{7em}{CaCu\textsubscript{3}Cr\textsubscript{2}Sb\textsubscript{2}O\textsubscript{12}} & \begin{tabular}{@{}l@{}}B-site rocksalt \cite{Byeon:2005}\end{tabular} &\multirow{2}{3.6em}{$1960,\ 26$} & $0.37$ & \multirow{2}{4.1em}{$0.60,\ 0.27$} \\
& \begin{tabular}{@{}l@{}}A-site body-centered \cite{Byeon:2005}\end{tabular} & & $0.00$ & \\
\hline
\end{tabular}
\begin{tablenotes}
\small
$N_{\mathrm{total}}$ is the number of all possible cation arrangements in a $2 \times 2 \times 2$ supercell, while $N_{\mathrm{unique}}$ is the number of symmetrically unique cation configurations. $P_{\mathrm{exp}}$ is the likelihood of forming the arrangement equivalent to experimental ordering (Eq. \ref{eqn:thermo_prob}). $S_{\mathrm{conf}}$ is the configurational entropy (Eq. \ref{eqn:conf_entropy}) with a range between $S_{\mathrm{residual}}$ and $1$, where $S_{\mathrm{residual}}$ is the residual entropy (Eq. \ref{eqn:descriptor_construction_residual_entropy}). Nd\textsubscript{2}CuSnO\textsubscript{6}, Pr\textsubscript{2}CuSnO\textsubscript{6}, and Sm\textsubscript{2}CuSnO\textsubscript{6} have only been synthesized under high pressure \cite{Azuma:1998}.
\end{tablenotes}
\end{threeparttable}
\end{table}

These physics-informed thermodynamic descriptors, e.g., $S_{\mathrm{conf}}$ and $P_{\mathrm{r}}$, can be generalized beyond the binary classification of A\textsubscript{2}B'B"O\textsubscript{6} into cation-ordered perovskites with rocksalt ordering or disordered ones (Table \ref{table:hold_out_test_set}). For the additional evaluation set of oxides with either B-site layered ordering \cite{Azuma:1998} or A-site columnar \cite{Aimi:2014} or body-centered ordering \cite{Yamada:2008,Long:2009}, to the best of our knowledge, there is a lack of atomic descriptors for rationalizing such exotic cation orderings. However, our thermodynamic descriptors suggest that the experimentally observed orderings for these oxides tend to be the ones with the highest thermodynamic likelihood of forming the configuration equivalent to experimental ordering ($P_{\mathrm{exp}}$). We also found there are negligible differences between $S_{\mathrm{conf}}$ and $S_{\mathrm{residual}}$, suggesting single dominant ordered phases, because a large difference would indicate that there are alternative orderings with a significant population to lead to disorder. These results highlight the power of $S_{\mathrm{conf}}$ and $P_{\mathrm{exp}}$ in predicting uncommon cation orderings. Nevertheless, extending the compositional space to perovskites with cation orderings in both A and B sites makes it difficult to find maximal and minimal values for $P_{\mathrm{exp}}$ and  $S_{\mathrm{conf}}$, respectively. For example, for CaCu\textsubscript{3}Cr\textsubscript{2}Sb\textsubscript{2}O\textsubscript{12} with experimentally observed B-site rocksalt and A-site body-centered orderings \cite{Byeon:2005}, the $P_{\mathrm{exp}}$ for such B-site and A-site orderings are $0.37$ and $0.00$, respectively. Moreover, the $S_{\mathrm{conf}}$ ($0.60$) of CaCu\textsubscript{3}Cr\textsubscript{2}Sb\textsubscript{2}O\textsubscript{12} is much higher than its $S_{\mathrm{residual}}$ ($0.27$). Such a mismatch between simulations and experiments can be potentially explained by the too-high configurational multiplicity of cation arrangements (i.e., $1960$ possible cation arrangements and $26$ symmetrically unique cation configurations in a $2 \times 2 \times 2$ supercell), as the energy gap between two symmetrically distinct configurations can be too small relative to the uncertainty in DFT (${\sim}20\ \mathrm{meV}$ per atom, \cite{Hautier:2012}). However, these energetically comparable configurations are hypothesized to have a minimal difference in their arrangements and properties. Thus, evaluating either of such arrangements might only result in a negligible influence on the utilization of cation ordering descriptors in investigating such complex perovskites. Overall, we confirmed the generalizability of cation ordering descriptors in understanding the various types of cation ordering across a broad compositional space.

\phantomsection
\addcontentsline{toc}{subsection}{Assessing the robustness and expressivity of cation ordering descriptors}
\subsection*{Assessing the robustness and expressivity of cation ordering descriptors}

\begin{figure}[hbt!]
\phantomsection
\begin{center}
\includegraphics[max size={\textwidth}{\textheight}]{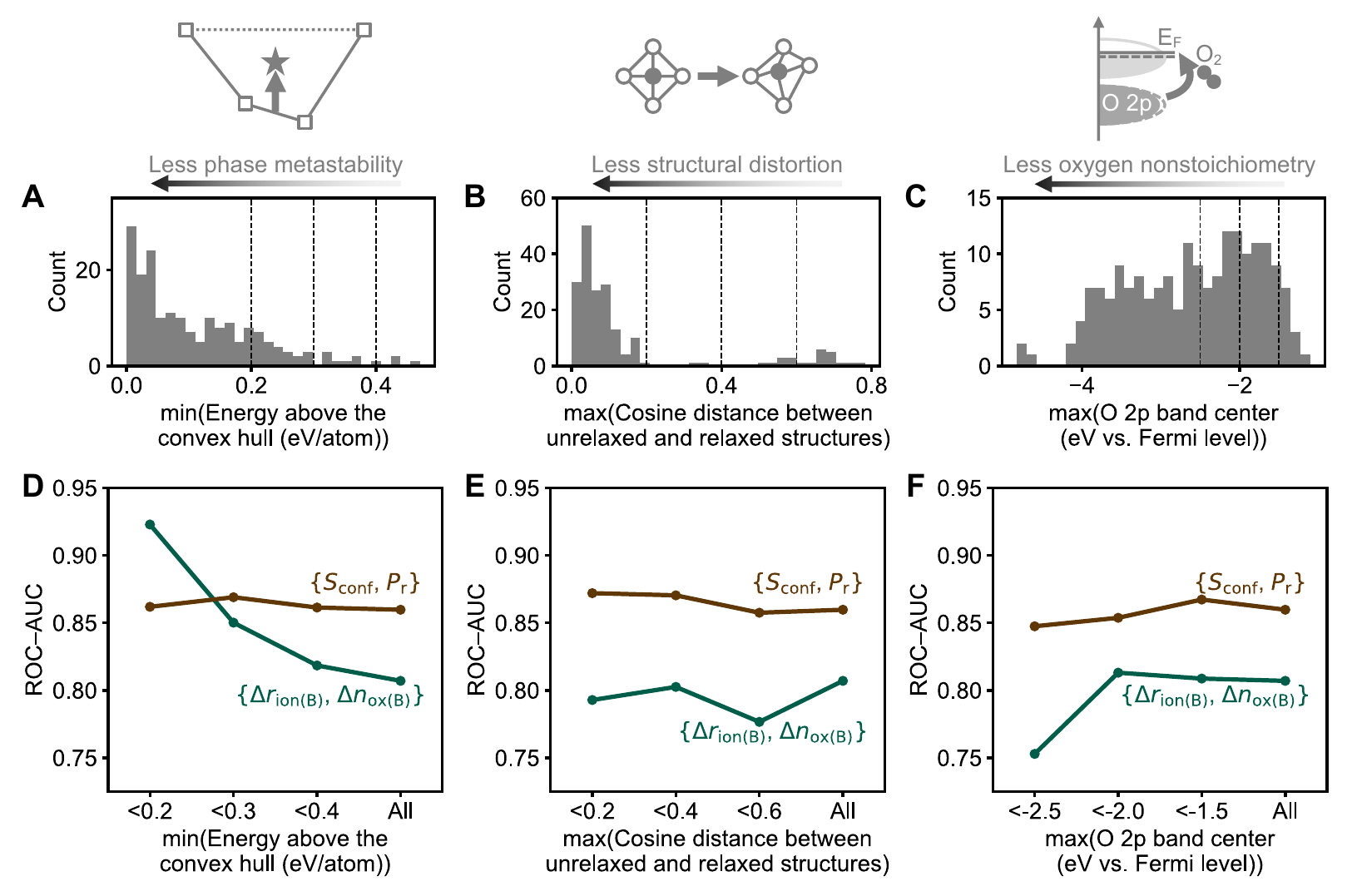}
\caption{
\textbf{Sensitivity analysis of thermodynamic and atomic descriptors.}
(\textbf{A} to \textbf{C}) Statistics of selected DFT-computed properties for the experimental dataset of $190$ A\textsubscript{2}B'B"O\textsubscript{6} oxides, including the minimum energy above the convex hull, the maximum cosine distance between the Matminer local order parameter fingerprint \cite{Ward:2018} of initialized and relaxed supercells, and the maximum O 2p band center relative to the Fermi level, which capture their tendency for phase metastability \cite{Sun:2016}, structural distortion \cite{Law:2023}, and oxygen nonstoichiometry \cite{Giordano:2022}, respectively. (\textbf{D} to \textbf{F}) ROC--AUC scores of top-performing 2D descriptors in classifying these perovskites as cation-ordered or disordered before and after filtering them by limiting the range of the examined properties. $\Delta r_{\mathrm{ion(B)}}$ and $\Delta n_{\mathrm{ox(B)}}$ are the differences between the ionic radii and oxidation states of B-site ions, respectively. $S_{\mathrm{conf}}$ and $P_{\mathrm{r}}$ are the configuration entropy (Eq. \ref{eqn:conf_entropy}) and the likelihood of generating the rocksalt configuration (Eq. \ref{eqn:thermo_prob}), respectively. The ROC curves of these two 2D descriptors are provided in Fig. \ref{fig:SI_sensitivity_analysis}.
}
\label{fig:Main_sensitivity_analysis}
\end{center}
\end{figure}

We verified that these thermodynamic descriptors are physically robust, and their performance does not obviously degrade as a function of variation in other features, providing a systematic, comprehensive benchmark between theory and experiments. The DFT-computed formation energetics of model cation arrangements for the $190$ A\textsubscript{2}B'B"O\textsubscript{6} perovskites are consistent with the fact that they have been experimentally synthesized. $81\%$ of the calculated oxides have energies above the convex hull ($E_\mathrm{hull}$) in their compositional phase diagrams below $0.2\ \mathrm{eV}$ per atom (Fig. \ref{fig:Main_sensitivity_analysis}A). This range agrees with the previous empirical limit of $E_\mathrm{hull}$ for generating metastable oxides \cite{Sun:2016}. Interestingly, such phase metastability constrains the performance of atomic descriptors. Filtering out the remaining $19\%$ perovskites with $E_\mathrm{hull} > 0.2\ \mathrm{eV}$  boosted the ROC--AUC of $\{\Delta r_{\mathrm{ion(B)}}, \Delta n_{\mathrm{ox(B)}}\}$ from $0.81$ to $0.92$ (Fig. \ref{fig:Main_sensitivity_analysis}D). However, no changes in the AUC were shown for thermodynamic descriptors such as $\{S_{\mathrm{conf}}, P_{\mathrm{r}}\}$. This difference can be possibly rationalized by a bias in dataset distribution when conceiving traditional atomic descriptors, where the focus might have been disproportionately placed on perovskites that are less metastable and more synthetically accessible. In contrast, the data-driven evaluation across a wide oxide space in this work gives rise to more heuristic-free, unbiased \cite{Peng2:2022} design principles of cation ordering, more independent of the phase metastability of the composition itself.

Similarly, other structural and compositional features, including structural distortion and oxygen nonstoichiometry, have been proposed to affect cation ordering in A\textsubscript{2}B'B"O\textsubscript{6} oxides \cite{King:2010}. We quantified the structural distortion in these oxides using the difference between cubic and DFT-relaxed supercells \cite{Ward:2018,Law:2023} (Fig. \ref{fig:Main_sensitivity_analysis}B) and estimated their tendency for showing oxygen nonstoichiometry using the O 2p band center relative to the Fermi level \cite{Giordano:2022} (Fig. \ref{fig:Main_sensitivity_analysis}E). Similar to $E_\mathrm{hull}$, for structural distortion and oxygen nonstoichiometry, $\{S_{\mathrm{conf}}, P_{\mathrm{r}}\}$ is also more unbiased than $\{\Delta r_{\mathrm{ion(B)}}, \Delta n_{\mathrm{ox(B)}}\}$ (Figs. \ref{fig:Main_sensitivity_analysis} and \ref{fig:SI_sensitivity_analysis}). Therefore, we established a co-validation between simulations and experiments and highlighted the robustness of thermodynamic descriptors for inferring cation ordering in these multicomponent perovskite oxides.

\begin{figure}[hbt!]
\phantomsection
\begin{center}
\includegraphics[max size={\textwidth}{\textheight}]{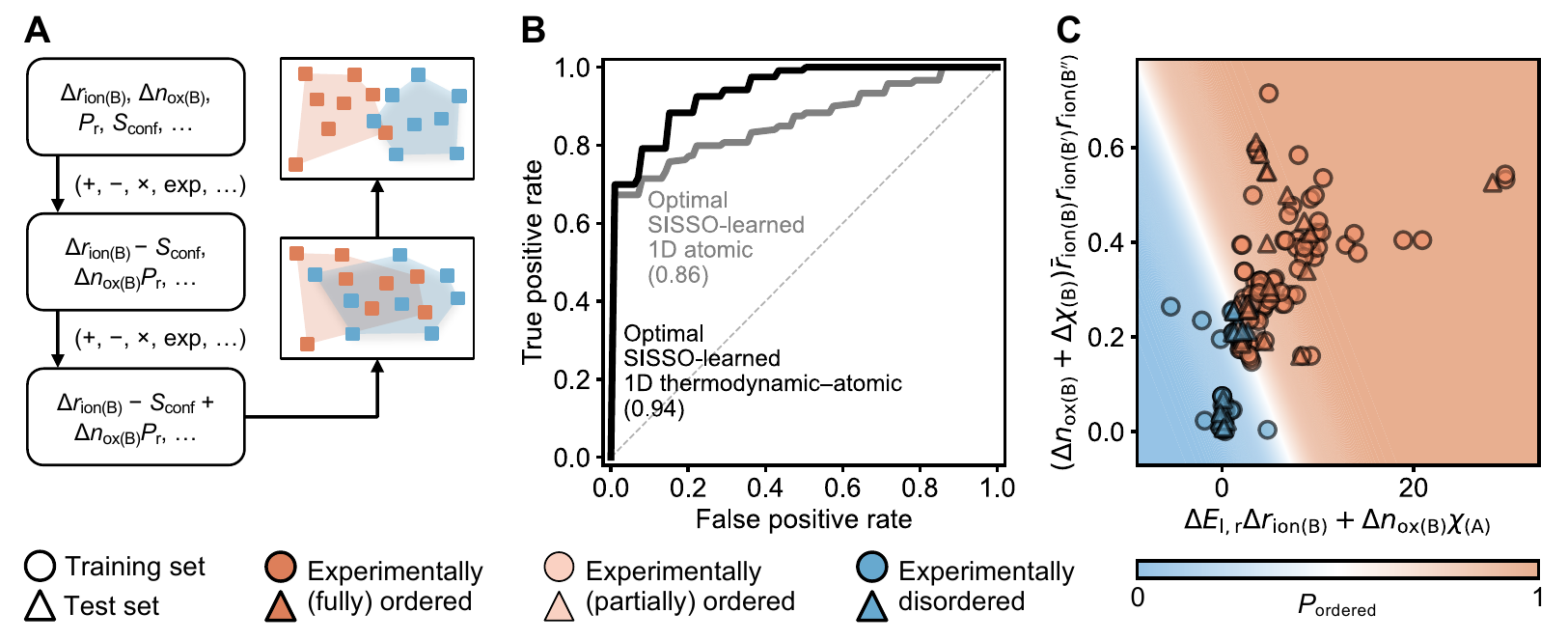}
\caption{
\textbf{SISSO-learned cation ordering descriptors.}
(\textbf{A}) Schematic illustration of the SISSO method, which searches for top-performing descriptors out of up to millions of candidates using a combinatorial set of primary features and mathematical operations. SISSO selects the optimal descriptors from these candidates by minimizing the domain overlap between cation-disordered and ordered perovskite oxides. (\textbf{B}) Optimal performance of the SISSO method using logistic regression with five-fold cross-validation, quantified by ROC curves with AUC in parentheses. Dashed lines show the ROC curves of a random classifier (AUC $= 0.5$). The AUC scores of the SISSO method with different settings are provided in Table \ref{table:SI_sisso_performance}. (\textbf{C}) Decision boundary of a 2D descriptor constructed using the best SISSO-learned 1D atomic and thermodynamic--atomic descriptors. $P_{\mathrm{ordered}}$ is the predicted likelihood that an oxide should be classified as cation-ordered through logistic regression. The definitions of all primary features used for the SISSO analysis are provided in Table \ref{table:SI_sisso_definition}.
}
\label{fig:Main_sisso_descriptors}
\end{center}
\end{figure}

We further benchmarked the expressivity of multi-dimensional thermodynamic descriptors and showed that the linear combination of physically interpretable parameters can well capture the cation ordering in multicomponent oxides. Specifically, we compared our top-performing descriptors, e.g., $\{S_{\mathrm{conf}}, P_{\mathrm{r}}, \Delta r_{\mathrm{ion(B)}}, \Delta n_{\mathrm{ox(B)}}\}$ and $\{\Delta E_{\mathrm{l,r}}, \Delta r_{\mathrm{ion(B)}}, \Delta n_{\mathrm{ox(B)}}\}$, with those built using the sure independence screening and sparsifying operator (SISSO) method \cite{Ouyang:2018}. SISSO examines up to millions of non-linear mathematical expressions from selected primary features (Table \ref{table:SI_sisso_definition}) and algebraic and functional operations, where the best descriptors can be further discovered (Fig. \ref{fig:Main_sisso_descriptors}A; Materials and Methods). Notably, adding non-linear expressivity gives rise to a moderate increase in descriptor performance for the binary classification of A\textsubscript{2}B'B"O\textsubscript{6} perovskites into cation-ordered or disordered ones (Fig. \ref{fig:Main_sisso_descriptors}B). SISSO-learned one-dimensional (1D) atomic and thermodynamic--atomic descriptors have $86\%$ and $94\%$ accuracies in ranking cation-ordered perovskites above their disordered counterparts, which are higher than those of linear models based on descriptors such as $\{S_{\mathrm{conf}}, P_{\mathrm{r}}\}$ ($81\%$) and $\{S_{\mathrm{conf}}, P_{\mathrm{r}}, \Delta r_{\mathrm{ion(B)}}, \Delta n_{\mathrm{ox(B)}}\}$ ($90\%$), respectively. Interestingly, the parameters in such simple linear descriptors are also the most important primary features in building the optimal SISSO-derived descriptors (Tables \ref{table:SI_sisso_importance_atomic} and \ref{table:SI_sisso_importance_thermo_atomic}), demonstrating that these key parameters capture cation ordering well. For instance, the best 1D thermodynamic--atomic and atomic descriptors are $\Delta E_{\mathrm{l,r}} \Delta r_{\mathrm{ion(B)}} + \Delta n_{\mathrm{ox(B)}} \chi_{\mathrm{(A)}}$ and $(\Delta n_{\mathrm{ox(B)}} + \Delta \chi_{\mathrm{(B)}}) \bar{r}_{\mathrm{ion(B)}} r_{\mathrm{ion(B')}} r_{\mathrm{ion(B'')}}$, respectively, which can be combined as a 2D descriptor that accurately distinguishes cation-ordered perovskites from the disordered ones (Fig. \ref{fig:Main_sisso_descriptors}C). In these SISSO-learned descriptors, apart from $\Delta r_{\mathrm{ion(B)}}$ and $\Delta n_{\mathrm{ox(B)}}$, the presence of other atomic parameters, including the electronegativity of the A-site cations ($\chi_{\mathrm{(A)}}$) and the average ($\bar{r}_{\mathrm{ion(B)}}$) and product between the ionic radii of the B-site cations ($r_{\mathrm{ion(B')}} r_{\mathrm{ion(B'')}}$), shows underexplored atomic features worthy of potential studies in future work. Overall, we systematically evaluated descriptors beyond the linear combination of atomic and thermodynamic parameters using the SISSO method, validating that linear descriptors (Figs. \ref{fig:Main_binary_classfication}, \ref{fig:SI_additional_descriptor_performance}, and \ref{fig:SI_filtering_partially_experimentally_ordered}) are physically expressive and can well characterize the cation ordering in multicomponent perovskite oxides.

\phantomsection
\addcontentsline{toc}{subsection}{Using cation ordering descriptors to boost chemical space exploration}
\subsection*{Using cation ordering descriptors to boost chemical space exploration}

Since the proposed descriptors can elucidate cation ordering, we evaluated how to use them for boosting the HTVS of multicomponent oxides. The focus was balancing the computational cost and property prediction accuracy over the vast configurational space of cation arrangements that will be responsible for the observed properties at any given composition. Instead of using massive supercells to characterize the quasirandom cation sublattices of multicomponent oxides\cite{Ferrari:2023,Jiang:2016}, HTVS can be conducted through DFT calculations of smaller supercells with model cation arrangements (Fig. \ref{fig:Main_dataset_statistics}) and quantify whether one ordered phase will dominate in structures and properties, or instead, an ensemble of disordered structures will exist, giving a weighted average of the properties over various possible cation configurations in the supercell.

\begin{figure}[hbt!]
\phantomsection
\begin{center}
\includegraphics[max size={\textwidth}{\textheight}]{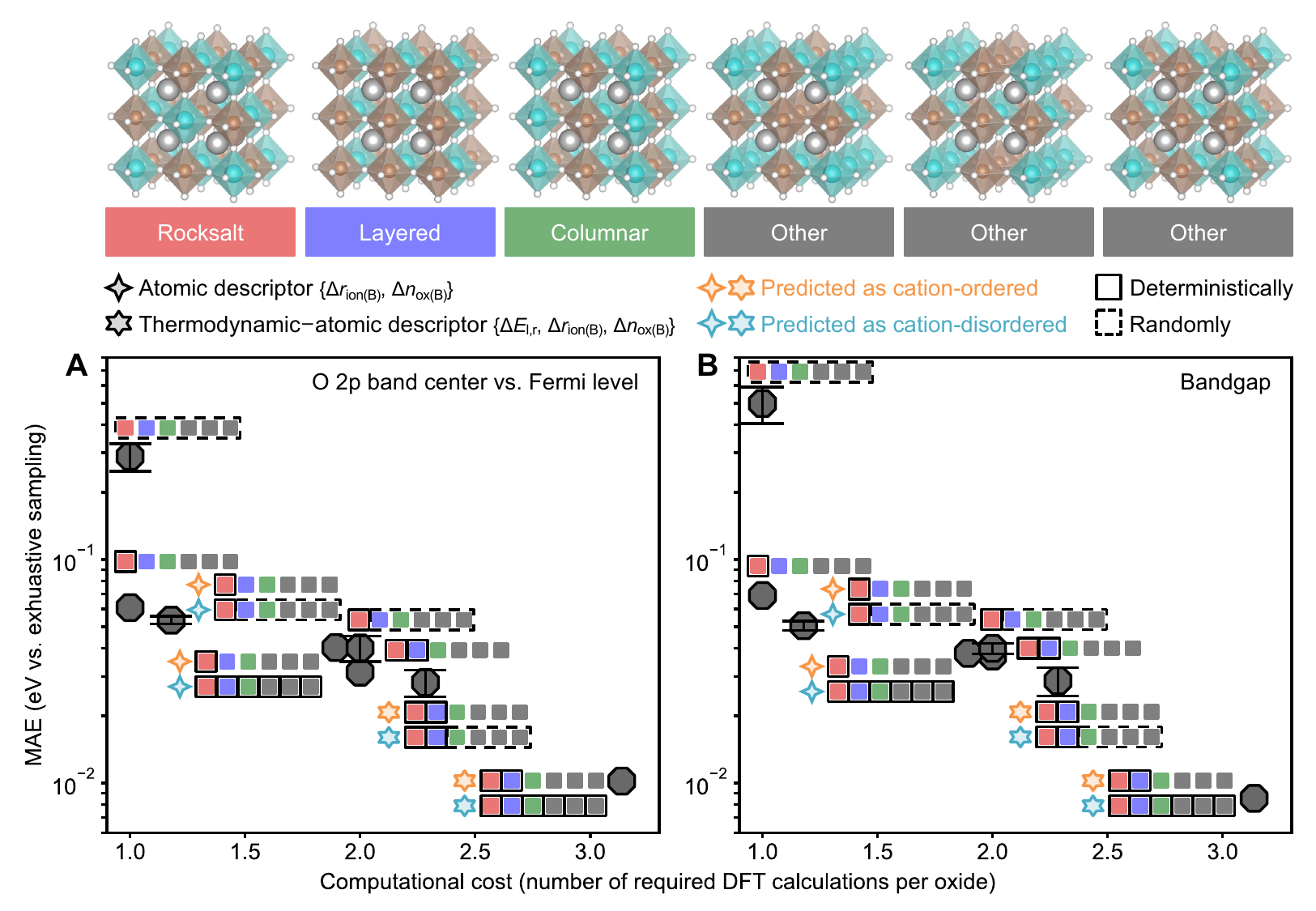}
\caption{
\textbf{Descriptor-accelerated chemical space exploration by balancing accuracy and cost.}
(\textbf{A} and \textbf{B}) Trade-offs between the accuracy and cost of various descriptor-driven and descriptor-free strategies for sampling the configurational space of cation arrangements in $190$ A\textsubscript{2}B'B"O\textsubscript{6} perovskite oxides. The O 2p band center and bandgap are the targeted key properties, with representative sampling results provided in Figs. \ref{fig:SI_selection_strategies_O2p} and \ref{fig:SI_selection_strategies_bandgap}, respectively. For each strategy, its computational cost was quantified based on the number of required DFT calculations per oxide, and its accuracy was obtained from the MAE of sampled properties with respect to the computationally expensive ground truth based on the ensemble average of all configurations. $\Delta r_{\mathrm{ion(B)}}$ and $\Delta n_{\mathrm{ox(B)}}$ are the differences between the ionic radii and oxidation states of B-site ions, respectively. $\Delta E_{\mathrm{l,r}}$ is the energy gap between the layer and rocksalt arrangements. Error bars represent the standard deviations of $50$ independent tests of random sampling.
}
\label{fig:Main_selection_strategies}
\end{center}
\end{figure}

We assessed the accuracy and cost of various strategies for selecting limited representative cation arrangements in A\textsubscript{2}B'B"O\textsubscript{6} perovskites, seeking to replicate the prediction of weighted properties across all orderings at a lower cost (Fig. \ref{fig:Main_selection_strategies}). We compared strategies such as selecting a single cation ordering or multiple tentative orderings. In addition, we evaluated whether our descriptors can triage oxides expected to be ordered---requiring fewer property evaluations---from those expected to be disordered, which may require more exhaustive calculations.
 
Specifically, we compared mean absolute errors (MAE) of predicting properties, computed from a weighted average of orderings chosen by various strategies, with respect to the weighted average of the exhaustive---but prohibitively expensive---DFT calculations of all configurations (Materials and Methods). Notably, we focused on two key properties of interest, including the O 2p band center and bandgap of perovskite oxides, which have been shown to govern their critical functions, e.g., catalytic activity \cite{Grimaud:2013,Hwang:2021}, electrochemical stability \cite{Kuznetsov:2020,Peng:2022}, optical absorption \cite{Nechache:2015}, intrinsic electrical conductivity \cite{Ma:2021}, and oxygen surface exchange kinetics \cite{Jacobs:2022}.  

Particularly, we used $\{\Delta r_{\mathrm{ion(B)}}, \Delta n_{\mathrm{ox(B)}}\}$ and $\{\Delta E_{\mathrm{l,r}}, \Delta r_{\mathrm{ion(B)}}, \Delta n_{\mathrm{ox(B)}}\}$ as representative atomic and thermodynamic--atomic descriptors, respectively, to decide which oxides are likely disordered and require more calculations. While $S_{\mathrm{conf}}$ and $P_{\mathrm{r}}$ are more expressive than $\Delta E_{\mathrm{l,r}}$, they require exhaustive DFT evaluation of all orderings. Therefore, we selected $\Delta E_{\mathrm{l,r}}$ as the thermodynamic parameter, alongside $\Delta r_{\mathrm{ion(B)}}$ and $\Delta n_{\mathrm{ox(B)}}$, which are computationally efficient, as these two parameters can be relatively easily deduced from perovskite compositions \cite{Bartel:2019}. 

Descriptor-driven selection strategies were found to representatively, yet efficiently explore the massive configurational space of cation ordering in multicomponent perovskite oxides. Interestingly, while randomly generating cation arrangements has been widely leveraged in literature \cite{Choubisa:2023,Tran:2023}, this strategy can lead to MAEs of $300\ \mathrm{meV}$ (Figs. \ref{fig:Main_selection_strategies}A and \ref{fig:SI_selection_strategies_O2p}) and $500\ \mathrm{meV}$ (Figs. \ref{fig:Main_selection_strategies}B and \ref{fig:SI_selection_strategies_bandgap}) for sampling the O 2p band center and bandgap of A\textsubscript{2}B'B"O\textsubscript{6} perovskite oxides, respectively, by overweighting the likelihood of unrealistic orderings. Unfortunately, an error of $300\ \mathrm{meV}$ in the O 2p band center of oxides can overestimate or underestimate their electrochemical oxygen evolution or surface exchange rates by one order of magnitude \cite{Jacobs:2019}. Similarly, a difference of $500\ \mathrm{meV}$ in the bandgap can modulate an oxide from being optimal to impractical for solar light capture and photoelectrochemical water splitting \cite{Castelli:2012}. In contrast to random sampling, manually predefining a cation arrangement for HTVS\cite{Ma:2021,Jacobs:2022} can give rise to a much lower error in estimating the cation ordering--averaged properties. For example, the MAEs can be decreased by one order of magnitude when focusing on the experimentally dominant configuration of cation-ordered perovskite oxides, such as the rocksalt arrangement for A\textsubscript{2}B'B"O\textsubscript{6} perovskites (Fig. \ref{fig:Main_selection_strategies}). If it is computationally affordable to examine more than one symmetrically inequivalent cation arrangement per oxide, cation ordering descriptors, e.g., $\{\Delta E_{\mathrm{l,r}}, \Delta r_{\mathrm{ion(B)}}, \Delta n_{\mathrm{ox(B)}}\}$, can be utilized to triage ordered perovskite oxides from disordered ones that need further calculations and decrease the MAEs by another order of magnitude to reach $10\ \mathrm{meV}$ (Fig. \ref{fig:Main_selection_strategies}A) and $9\ \mathrm{meV}$ (Fig. \ref{fig:Main_selection_strategies}B) for sampling the O 2p band center and bandgap, respectively, with an average cost of three DFT calculations per oxide. Such descriptor-driven strategies reside at the Pareto front of accuracy--cost trade-offs, highlighting their crucial roles in accelerating HTVS by predicting which perovskites should be cation-disordered and require more extensive DFT exploration of cation configurations than their ordered counterparts.

\phantomsection
\addcontentsline{toc}{section}{Discussion}
\section*{Discussion}

In summary, we utilized high-throughput atomistic simulations and machine learning to build physically interpretable design principles that universally rationalize and accurately predict the experimental cation ordering in multicomponent perovskite oxides. Leveraging a dataset of $190$ experimentally studied A\textsubscript{2}B'B"O\textsubscript{6} double perovskites, we demonstrated that such ordering descriptors can be obtained from first-principles calculations of model cation arrangements in computationally inexpensive $2 \times 2 \times 2$ supercells. These descriptors can be rationalized as physics-informed parameters such as thermodynamic likelihood and configurational entropy, outperforming their traditional atomic counterparts in classifying oxides as cation-ordered or disordered. Combining such thermodynamic and atomic parameters further leads to hybrid descriptors that can correctly rank up to $93\%$ perovskite compositions between cation-ordered and disordered, offering an experiment-backed, systematic validation of atomistic simulations and data science. We verified that these design principles are physically robust and expressive and can be applied to other multicomponent perovskite oxides with more exotic compositions and orderings. Lastly, we highlighted that such physics-driven descriptors can be leveraged to boost the high-throughput exploration of multicomponent oxides across a vast compositional space by facilitating an optimal balance between the breadth and cost of computationally expensive DFT sampling for different cation arrangements in these complex oxides.

The design principles of cation ordering established in this work can be potentially extended for rationalizing and inferring the chemical and structural orders in other multicomponent oxide chemistries. High-entropy oxides represent an exciting, yet largely uncharted chemical space, which creates exotic physical and chemical properties and flips traditional research paradigms by introducing profound configurational disorder in these new materials \cite{Aamlid:2023}. While these high-entropy oxides generally do not have long-range ordering, they can still exhibit local clustering \cite{Jiang:2021}, short-range ordering \cite{Lun:2021}, and site selectivity \cite{Johnstone:2022}. Unfortunately, most previous studies on high-entropy oxides have so far overlooked such local chemical and structural orders and the unique role of these local orders in influencing key material properties \cite{Aamlid:2023}. This limitation arises from the difficulty of quantitatively characterizing such chemical and structural orders, as their spatial scales evade most available experimental techniques \cite{Ferrari:2023}. To conquer this task, it will be advantageous to leverage simulations, which, in theory, have full access to determining a realistic atom-by-atom picture of these high-entropy oxides but, in practice, face the challenge of accurately and efficiently sampling their high-dimensional configurational spaces. This work can serve as a blueprint for future studies to elucidate the correlated arrangements of cations in high-entropy oxides and predict their degree of order or disorder at the atomistic level. From a broad perspective, we aim to couple physics-informed principles with data-driven tools in future work to simultaneously navigate the compositional and configurational spaces of high-entropy materials and develop effective design principles to accelerate their global optimization.

\phantomsection
\addcontentsline{toc}{section}{Materials and Methods}
\section*{Materials and Methods}

\phantomsection
\addcontentsline{toc}{subsection}{Experimental datasets}
\subsection*{Dataset curation}

A dataset of $190$ experimentally examined A\textsubscript{2}B'B"O\textsubscript{6} compositions was curated according to a recent review \cite{Vasala:2015}. We focused on oxides that are synthesizable under ambient pressure, where all but one known cation-ordered A\textsubscript{2}B'B"O\textsubscript{6} perovskites have B-site sublattices with rocksalt ordering \cite{Anderson:1993,Vasala:2015}. This only composition with B-site layered ordering under ambient pressure was not included for preventing too-massive data imbalance in training and assessing machine learning models. Overall, this dataset has $119$ cation-ordered oxides with rocksalt ordering and $71$ cation-disordered oxides. Among these $119$ perovskite oxides, eight of them only have partially ordered cation arrangements, as shown by their low ($<0.5$) experimentally measured long-range order parameters \cite{Vasala:2015}. Apart from this dataset, an additional evaluation set with less common compositions and orderings was built according to experimental literature \cite{Azuma:1998,Aimi:2014,Yamada:2008,Long:2009,Byeon:2005}.

\phantomsection
\addcontentsline{toc}{subsection}{DFT calculations}
\subsection*{DFT calculations}

Periodic plane-wave spin-polarized DFT calculations were performed to assess the electronic structures and energetics of multicomponent perovskite oxides and their cation ordering. We used Perdew--Burke--Ernzerhof (PBE) functional \cite{Perdew:1996} as implemented in the Vienna Ab initio Simulation Package \cite{Kresse:1993,Kresse:1996} and projector augmented wave method \cite{Blochl:1994} for the description of core-electron interaction, and the plane-wave cutoff was set to $520\ \mathrm{eV}$. DFT calculations were conducted with Hubbard $U$ correction for V 3d ($U_{\mathrm{V}} = 3.25\ \mathrm{eV}$), Cr 3d ($U_{\mathrm{Cr}} = 3.7\ \mathrm{eV}$), Mn 3d ($U_{\mathrm{Mn}} = 3.9\ \mathrm{eV}$), Fe 3d ($U_{\mathrm{Fe}} = 5.3\ \mathrm{eV}$), Co 3d ($U_{\mathrm{Co}} = 3.32\ \mathrm{eV}$), Ni 3d ($U_{\mathrm{Ni}} = 6.2\ \mathrm{eV}$), Mo 4d ($U_{\mathrm{Mo}} = 4.38\ \mathrm{eV}$), and W 5d ($U_{\mathrm{W}} = 6.2\ \mathrm{eV}$) electrons, where the $U$ values were optimized by fitting the experimental formation enthalpies of binary oxides \cite{Wang:2006}. DFT calculations were initialized with high-spin ferromagnetic states in order to use a consistent and tractable set of magnetic structures. These computational settings are fully compatible with the Materials Project database \cite{Jain:2013}, as the Materials Project represents a good standard for high-throughput materials discoveries, and we can utilize this expansive database to derive specific properties of multicomponent perovskite oxides (e.g., $E_\mathrm{hull}$). Utilizing more advanced density functional and more diverse initial magnetic configurations might improve the accuracy of DFT calculations but are too computationally expensive for high-throughput exploration. Therefore, in this work, we focused on PBE-level calculations with high-spin ferromagnetic initialization and showed that they could well capture experimental ordering across a wide oxide space.

We used an in-house automated high-throughput DFT pipeline for structure optimization and electronic structure calculation. Initial $2 \times 2 \times 2$ cubic supercells were built by generating all symmetrically distinct cation arrangements with the Atomic Simulation Environment \cite{Larsen:2017} and Python Materials Genomics (pymatgen) \cite{Ong:2013}. The initialized lattice parameters were set to be $8\ \mathrm{\mathring{A}}$, and all atoms in these initialized cubic supercells could be displaced by a distance randomly sampled from a uniform distribution with a range between $0.01\ \mathrm{\mathring{A}}$ and $ 0.1\ \mathrm{\mathring{A}}$ to break symmetry and boost convergence. The convergence threshold for electronic steps was $10^{-6}\ \mathrm{eV}$ per unit cell, and the residual forces on all atoms were lower than $10^{-2}\ \mathrm{eV\ \mathring{A}}$. O 2p band centers were determined by taking the centroid of the O 2p projected density of states relative to the Fermi level. $E_\mathrm{hull}$ was computed using the pymatgen\cite{Ong:2013} by combining the DFT energies of multicomponent oxides with all data entries within the examined compositional space from the Materials Project \cite{Jain:2013}. The degree of structural distortion before and after DFT relaxation was estimated by the cosine distances between the Matminer local order parameter fingerprint \cite{Ward:2018} of initialized and relaxed supercells based on a method developed in a recent work \cite{Law:2023}.

\phantomsection
\addcontentsline{toc}{subsection}{Descriptor construction}
\subsection*{Descriptor construction}

Cation ordering descriptors were constructed using material parameters, i.e., $\Delta E_{\mathrm{l,r}}$, $S_{\mathrm{conf}}$, $P_{\mathrm{r}}$, $\Delta \chi_{\mathrm{(B)}}$, $\Delta r_{\mathrm{ion(B)}}$, and $\Delta n_{\mathrm{ox(B)}}$, where these parameters were used as individual 1D descriptors and combined as multi-dimensional descriptors. First, $P_{\mathrm{r}}$ and $S_{\mathrm{conf}}$ were derived from the DFT-computed energies of all symmetrically distinct cation arrangements based on Eqs. \ref{eqn:thermo_prob} and \ref{eqn:conf_entropy}, respectively. We set $T = 1300\ \mathrm{K}$ based on an extensive survey on the synthesis conditions of perovskite oxides \cite{Karpovich:2023}, where long-term calcination at comparable temperatures followed by rapid quenching has been widely used in experimental literature to kinetically trap these oxides at their high-temperature, thermodynamically favorable perovskite phases \cite{Aamlid:2023,Martinolich:2017}. Although based on convention \cite{Sethna:2021}, the configurational entropy of cation arrangements should be:
\begin{equation}
\label{eqn:descriptor_construction_unnormalized_conf_entropy}
\hat{S}_{\mathrm{conf}} = -\sum_{i=1}^n \Omega_i \frac{\exp{\left( -\frac{E_i}{k_{\mathrm{B}}T} \right)}}{\sum_{j=1}^{N_{\mathrm{unique}}} \Omega_j \exp{\left( -\frac{E_j}{k_{\mathrm{B}}T} \right)}} \ln{ \frac{\exp{\left( -\frac{E_i}{k_{\mathrm{B}}T} \right)}}{\sum_{j=1}^{N_{\mathrm{unique}}} \Omega_j \exp{\left( -\frac{E_j}{k_{\mathrm{B}}T} \right)}} }
\end{equation}
the range of $\hat{S}_{\mathrm{conf}}$ depends too strongly on the degeneracy. Specifically, if the energies of all cation arrangements are equal, $\hat{S}_{\mathrm{conf}}$ is maximized as $\ln{N_{\mathrm{total}}}$, where $N_{\mathrm{total}}$ is the total number of all possible cation configurations ($N_{\mathrm{total}} = 70$ for A\textsubscript{2}B'B"O\textsubscript{6} in Fig. \ref{fig:Main_dataset_statistics}):
\begin{equation}
\label{eqn:descriptor_construction_n_total}
N_{\mathrm{total}} = \sum_{k=1}^{N_{\mathrm{unique}}} \Omega_k
\end{equation}
On the other hand, if a cation arrangement $\alpha$ is much more thermodynamically favorable than its symmetrically inequivalent counterparts, $\hat{S}_{\mathrm{conf}}$ is minimized as $\ln{\Omega_{\alpha}}$. To facilitate the intuitive comparison of the configurational entropy, we further normalized it to make it dimensionless:
\begin{equation}
\label{eqn:descriptor_construction_normalized_conf_entropy}
\begin{gathered}
S_{\mathrm{conf}} = \frac{\hat{S}_{\mathrm{conf}}}{\ln{N_{\mathrm{total}}}} = \frac{\hat{S}_{\mathrm{conf}}}{\ln{\sum_{k=1}^{N_{\mathrm{unique}}} \Omega_k}} \\
= -\sum_{i=1}^n \frac{\Omega_i}{\ln{\sum_{k=1}^{N_{\mathrm{unique}}} \Omega_k}} \frac{\exp{\left( -\frac{E_i}{k_{\mathrm{B}}T} \right)}}{\sum_{j=1}^{N_{\mathrm{unique}}} \Omega_j \exp{\left( -\frac{E_j}{k_{\mathrm{B}}T} \right)}} \ln{ \frac{\exp{\left( -\frac{E_i}{k_{\mathrm{B}}T} \right)}}{\sum_{j=1}^{N_{\mathrm{unique}}} \Omega_j \exp{\left( -\frac{E_j}{k_{\mathrm{B}}T} \right)}} }
\end{gathered}
\end{equation}
with a range between $S_{\mathrm{residual}}$ and $1$, where $S_{\mathrm{residual}}$ is the residual entropy \cite{Sethna:2021} originating from the configurational multiplicity of the symmetrically distinct ground-state arrangement $\alpha$:
\begin{equation}
\label{eqn:descriptor_construction_residual_entropy}
 S_{\mathrm{residual}} = \frac{\ln{\Omega_{\alpha}}}{\ln{N_{\mathrm{total}}}} = \frac{\ln{\Omega_{\alpha}}}{\ln{\sum_{k=1}^{N_{\mathrm{unique}}} \Omega_k}}
\end{equation}
Second, $\Delta E_{\mathrm{l,r}}$ was calculated based on the formation energies of the layer ($E_{\mathrm{layered}}$) and rocksalt cation arrangements ($E_{\mathrm{rocksalt}}$):
\begin{equation}
\label{eqn:descriptor_construction_energy_gap}
\Delta E_{\mathrm{l,r}} = \frac{E_{\mathrm{layered}} - E_{\mathrm{rocksalt}}}{k_{\mathrm{B}}T}
\end{equation}
where the energy is normalized by $k_{\mathrm{B}}T$ to make it dimensionless. Lastly, $\Delta \chi_{\mathrm{(B)}}$, $\Delta r_{\mathrm{ion(B)}}$, and $\Delta n_{\mathrm{ox(B)}}$ were estimated by calculating the differences between the Pauling electronegativities \cite{Allred:1961}, Shannon effective ionic radii \cite{Shannon:1976}, and nominal oxidation states of the two different B-site cations in double perovskites, respectively. Similar to $\Delta E_{\mathrm{l,r}}$, $\Delta r_{\mathrm{ion(B)}}$ is normalized by the ionic radius of an oxygen anion ($r_{\mathrm{ion(O^{2-})}} = 1.40\ \mathrm{\mathring{A}}$) to make it dimensionless.

\phantomsection
\addcontentsline{toc}{subsection}{Logistic regression}
\subsection*{Logistic regression}

A logistic regression model was utilized to assess the performance of descriptors in classifying perovskite oxides into cation-ordered and disordered structures. Material parameters, including $\Delta E_{\mathrm{l,r}}$, $S_{\mathrm{conf}}$, $P_{\mathrm{r}}$, $\Delta \chi_{\mathrm{(B)}}$, $\Delta r_{\mathrm{ion(B)}}$, and $\Delta n_{\mathrm{ox(B)}}$, were used to construct one- or multi-dimensional descriptors as input features. For an input feature $(x_1,\ldots,x_M)$, where $M$ is its dimension, the model predicts the likelihood that an oxide should be classified as cation-ordered:
\begin{equation}
\label{eqn:logistic_regression_likelihood}
P_{\mathrm{ordered}} = \frac{1}{1 + \exp(-\sum_{i=1}^M \theta_i x_i - \theta_0)}
\end{equation}
where $\theta_0,\ldots,\theta_M$ are learnable parameters. An oxide is predicted as cation-ordered if $P_{\mathrm{ordered}} > 0.5$ and disordered if $P_{\mathrm{ordered}} < 0.5$. The logistic regression model was trained by minimizing its loss function in the form of negative log-likelihood (NLL) with $L_2$ regularization:
\begin{equation}
\label{eqn:logistic_regression_likelihood_loss}
l_{\mathrm{NLL}} = \lambda \sum_{i=0}^M \theta_i^2 - \sum_{j=1}^{N_{\mathrm{oxide}}} (y_{\mathrm{ordered},j} \ln{P_{\mathrm{ordered},j}} + (1 - y_{\mathrm{ordered},j}) \ln(1 - P_{\mathrm{ordered},j}))
\end{equation}
where $\lambda$ is the regularization hyperparameter, $N_{\mathrm{oxide}}$ is the number of oxides, $y_{\mathrm{ordered},j}$ is the label indicating whether an oxide $j$ is experimentally cation-ordered ($y_{\mathrm{ordered},j} = 1$ if ordered and $y_{\mathrm{ordered},j} = 0$ if disordered), and $P_{\mathrm{ordered},j}$ is the predicted likelihood if such an oxide should be classified as cation-ordered through logistic regression (Eq. \ref{eqn:logistic_regression_likelihood}). 

The model was trained using the scikit-learn \cite{Pedregosa:2011} and evaluated by both ROC–AUC scores and confusion matrices. Five-fold stratified cross-validation was utilized to rigorously quantify the accuracy and generalizability of model performance for predicting the cation ordering of unseen multicomponent perovskite oxides, given the random splits and the unbalanced numbers of experimentally cation-ordered and disordered oxides. The logistic regression model was first trained by combining four splits ($80\%$) and then tested on the one hold-out split ($20\%$). All input features were normalized by the mean and standard deviation of training sets. This process was iterated throughout the five folds, where only the $P_{\mathrm{ordered}}$ of test sets was collected for further assessment. ROC curves and AUC scores were obtained by averaging over the five folds, while the confusion matrices were based on the combined test set performance of these five folds. The ROC curves were obtained from the true positive rate (TPR) and false positive rate (FPR) for classifying perovskites as cation-ordered and disordered at various decision thresholds. The TPR is the percentage of experimentally cation-ordered oxides that are correctly predicted as ordered, while the FPR is the percentage of experimentally cation-disordered oxides that are incorrectly predicted as ordered. A perfect classifier maximizes the TPR as $1$ and minimizes the FPR as $0$, leading to an AUC of $1$. In contrast, a trivial classifier that infers cation ordering by randomly sampling a Bernoulli distribution based on the ratio of cation-ordered and disordered perovskites should always have equal TPR and FPR, resulting in an AUC of $0.5$.

\phantomsection
\addcontentsline{toc}{subsection}{SISSO}
\subsection*{SISSO}

We additionally leveraged the SISSO method \cite{Ouyang:2018} to assess the space of descriptors beyond the linear combination of atomic and thermodynamic parameters. Using an iterative approach, SISSO constructs non-linear expressions based on selected primary features and mathematical operations. These features were hypothesized to be potentially relevant for describing the cation ordering in multicomponent perovskite oxides and thus used as a starting point to build complex descriptors. The examined primary features are $\Delta E_{\mathrm{l,r}}$, $S_{\mathrm{conf}}$, $P_{\mathrm{r}}$, $\Delta \chi_{\mathrm{(B)}}$, $\Delta r_{\mathrm{ion(B)}}$, and $\Delta n_{\mathrm{ox(B)}}$, as well as nine additional atomic parameters, including the electronegativity, ionic radius, and oxidation state of the A-site cations and the average and product between the electronegativities, ionic radii, and oxidation states of the B-site cations (Table \ref{table:SI_sisso_definition}). These combinations of B-site atomic parameters were used such that the constructed descriptors are independent of the order of B' and B'' cations in A\textsubscript{2}B'B"O\textsubscript{6} perovskites. All such primary features are dimensionless, and thus they can be combined with a set of selected algebraic and functional operations recursively to expand the descriptor space beyond linearity. The assessed mathematical operations include $+$, $-$, $\times$, $\div$, $|\ |$, ${}^{-1}$, ${}^2$, ${}^3$, ${}^4$, ${}^{1/2}$ ${}^{1/3}$, ${}^{1/4}$, $\exp$, $\log$, $\sin$, and $\cos$. In total, we conducted two rounds of operations, leading to $498420$ candidates from only atomic features and $996440$ candidates from atomic and thermodynamic features after excluding numerically unstable ones (e.g., infinity).

Utilizing these features and operations, SISSO downselected the optimal descriptors from up to millions of candidates (Fig. \ref{fig:Main_sisso_descriptors}A). We used sure independence screening \cite{Ouyang:2018} to identify $1000$ top-performing non-linear expressions that minimize the domain overlap between cation-disordered and ordered perovskite oxides. Then, we utilized the sparsifying operator approach to further select the optimal SISSO-derived descriptors, where we experimented with two types of previous implementations based on convex hull \cite{Ouyang:2018} or a decision tree with a depth of $2$ \cite{Bartel:2019}, respectively. We assessed both implementations with and without using balanced class weights. Thus, when accounting for all settings, we carried out, in total, $16$ SISSO experiments (Table \ref{table:SI_sisso_performance}). As all such SISSO experiments were conducted with five-fold cross-validation, we focused on the AUC of the hold-out test splits in order to evaluate the generalizability of SISSO-learned descriptors for unseen perovskites. Overall, we found that using balanced class weights gave rise to better performance. Moreover, the convex hull method worked better for building atomic descriptors, while the decision tree method performed better in finding thermodynamic--atomic descriptors. We summarized the best SISSO performance of Table \ref{table:SI_sisso_performance} in Fig. \ref{fig:Main_sisso_descriptors}.

We found $\Delta E_{\mathrm{l,r}} \Delta r_{\mathrm{ion(B)}} + \Delta n_{\mathrm{ox(B)}} \chi_{\mathrm{(A)}}$ and $(\Delta n_{\mathrm{ox(B)}} + \Delta \chi_{\mathrm{(B)}}) \bar{r}_{\mathrm{ion(B)}} r_{\mathrm{ion(B')}} r_{\mathrm{ion(B'')}}$ to be the best 1D SISSO-learned descriptors in four out of five folds of cross-validation when selecting the optimal thermodynamic--atomic and atomic descriptors, respectively. This result shows that the downselection of such two 1D descriptors is relatively independent of the choices of training and test sets in cross-validation. We also examined the performance of the SISSO method with increased dimensionality using the residuals (i.e., those misclassified by the best 1D descriptors) in the search for a second dimension \cite{Bartel:2019}. Unfortunately, this second dimension was found to lead to negligible improvement in classification performance compared with the corresponding 1D descriptor (Table \ref{table:SI_sisso_performance}). Thus, we only focused on the best 1D descriptors in Fig. \ref{fig:Main_sisso_descriptors}.

Lastly, we note that, while we used SISSO to explore a space of up to millions of descriptor candidates, the potential for overfitting is minimal, as for 1D or 2D SISSO-derived descriptors, the SISSO method has only one or two independent learnable parameters that define the decision boundary, respectively. Moreover, finding different descriptors in five-fold cross-validation does not imply overfitting, as the best expressions from SISSO experiments reflect the intrinsically approximate nature of these expressions and also the unavoidable correlations between various descriptor candidates. In addition, another possible source of overfitting can come from a too-large dimensionality of SISSO-learned descriptors. However, this risk of overfitting is minimal in this work since we only examined 1D and 2D descriptors using the SISSO method. A more detailed discussion on these points was well presented in a previous study \cite{Bartel:2019}.

\phantomsection
\addcontentsline{toc}{subsection}{Chemical space exploration}
\subsection*{Chemical space exploration}

We tested various descriptor-driven and descriptor-free strategies to sample the configurational space of cation arrangements in $190$ A\textsubscript{2}B'B"O\textsubscript{6} perovskites and quantify the accuracy and cost of such strategies for the HTVS of multicomponent oxides. These strategies are composed of a set of potential actions, including sampling a specific symmetrically unique cation configuration deterministically, selecting a configuration randomly from a pool of cation arrangements, and leveraging descriptors to infer cation ordering and decide sampling methods accordingly. The computational cost was quantified based on the number of required DFT calculations per oxide. The accuracy was obtained from the MAE of the sampled properties ($\Phi_{\mathrm{sampled}}$) with respect to the computational ground truth based on the ensemble average of all possible configurations:
\begin{equation}
\label{eqn:chemical_space_exploration_ground_truth}
\Phi_{\mathrm{true}} = \sum_{i=1}^{N_{\mathrm{unique}}} \Phi_{i} P_i = \frac{\sum_{i=1}^{N_{\mathrm{unique}}} \Phi_{i} \Omega_i \exp{\left( -\frac{E_i}{k_{\mathrm{B}}T} \right)}}{\sum_{j=1}^{N_{\mathrm{unique}}} \Omega_j \exp{\left( -\frac{E_j}{k_{\mathrm{B}}T} \right)}}
\end{equation}
where $\Phi_{i}$ is the property (e.g., the O 2p band center or bandgap) of a symmetrically distinct configuration $i$. When selecting a configuration randomly from a number of candidates, $\Phi_{\mathrm{sampled}}$ was computed by assuming all other unsampled cation arrangements have the same $E_i$ and $\Phi_{i}$ as this sampled arrangement. When sampling both the rocksalt and layered configurations, all unsampled arrangements were assumed to have the same $E_i$ and $\Phi_{i}$ as the layered arrangement. For descriptor-driven strategies, five-fold stratified cross-validation was leveraged to examine the generalizability of descriptors for unseen perovskites by training the model on four splits and utilizing the prediction for the hold-out split. For strategies involving random sampling, $50$ independent tests were conducted to quantify their statistically averaged performance.

\phantomsection

\phantomsection
\addcontentsline{toc}{section}{Acknowledgments}
\bigskip\noindent{\bf\large Acknowledgments}

\noindent
\textbf{Funding:} This project was funded by the Advanced Research Projects Agency–Energy (ARPA-E), U.S. Department of Energy, under award number DE-AR0001220. This work used Expanse at San Diego Supercomputer Center through allocation DMR200068 from the Advanced Cyberinfrastructure Coordination Ecosystem: Services \& Support (ACCESS) program, which is supported by National Science Foundation grants 2138259, 2138286, 2138307, 2137603, and 2138296. This research used resources of the National Energy Research Scientific Computing Center (NERSC), a U.S. Department of Energy Office of Science User Facility located at Lawrence Berkeley National Laboratory, operated under Contract DE-AC02-05CH11231 using NERSC award m4074. We acknowledge the MIT Engaging cluster at the Massachusetts Green High-Performance Computing Center (MGHPCC) for providing high-performance computing resources. \textbf{Author contributions:} J.P. and R.G.-B. conceived the original idea. J.P. designed the studies. J.P. and J.D. performed the studies and analyzed the results. J.P. drafted the manuscript. All authors edited the manuscript. R.G.-B. supervised the project. \textbf{Competing interests:} The authors declare no competing interests. \textbf{Data and materials availability:} A repository for reproducing this work is available at https://github.com/learningmatter-mit/Perovskite-Ordering-Descriptors.

\bigskip\noindent{\bf\large List of supplementary materials}

\noindent
Figs. S1 to S7\\
Tables S1 to S4

\clearpage
\input{SI.tex}

\end{document}

%% file: SI.tex
\clearpage
\baselineskip18pt

\setcounter{page}{1}

\stepcounter{myequation}
\renewcommand{\theequation}{S\arabic{equation}}

\begin{center}

\vspace{36pt}{\Large Supplementary Materials for}\\[12pt]

{\bf\large Data-Driven, Physics-Informed Descriptors of Cation Ordering in Multicomponent Oxides}

\vspace{6pt} Jiayu Peng, James Damewood, Rafael Gómez-Bombarelli$^{*}$

\vspace{6pt}$^*$Corresponding author. Email: rafagb@mit.edu (R.G.-B.)

\end{center}

\vspace{5mm}

\noindent{\bf This PDF file includes:}


\vspace{-3pt}Figs. S1 to S7

\vspace{-3pt}Tables S1 to S4

\pagebreak

\clearpage
\newcounter{sfigure}
\renewcommand{\figurename}{Fig.}
\renewcommand{\thefigure}{S\arabic{sfigure}}

\addtocounter{sfigure}{1}
\begin{figure}
\phantomsection
\begin{center}
\includegraphics[max size={\textwidth}{\textheight},scale=0.95]{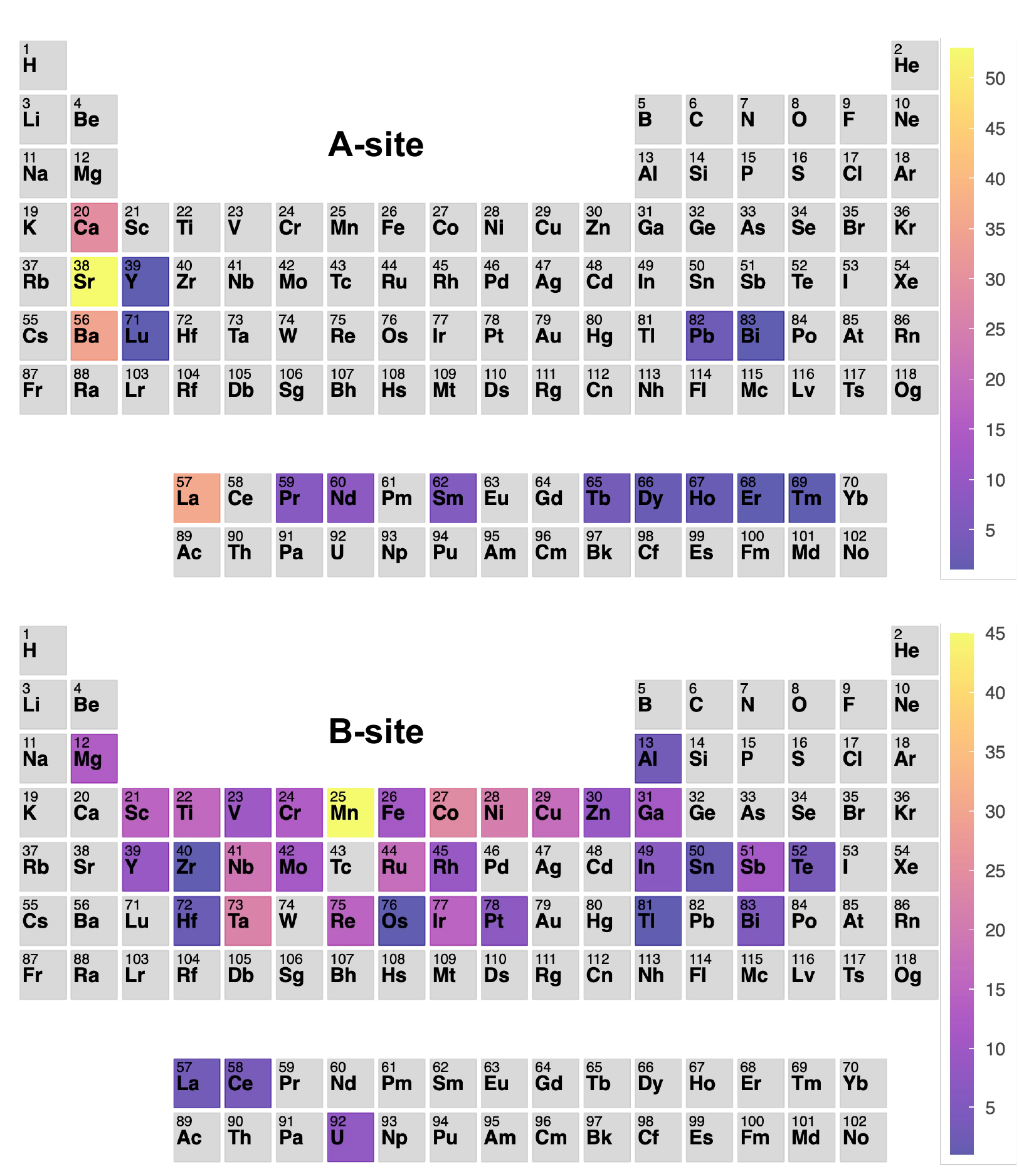}
\caption{
\textbf{Dataset statistics.}
Heat map of the occurrences of elements in A-site and B-site cations for the experimental dataset of $190$ A\textsubscript{2}B'B"O\textsubscript{6} double perovskites. These cations include a wide variety of elements that cover (but are not limited to) commonly seen A-site (alkaline earth and rare earth metals) and B-site ions (transition metals) in perovskite oxides, highlighting that this dataset contains oxide compositions across a broad space of multicomponent perovskites.
}
\label{fig:SI_elemental_info}
\end{center}
\end{figure}

\addtocounter{sfigure}{1}
\begin{figure}
\phantomsection
\begin{center}
\includegraphics[max size={\textwidth}{\textheight},scale=0.75]{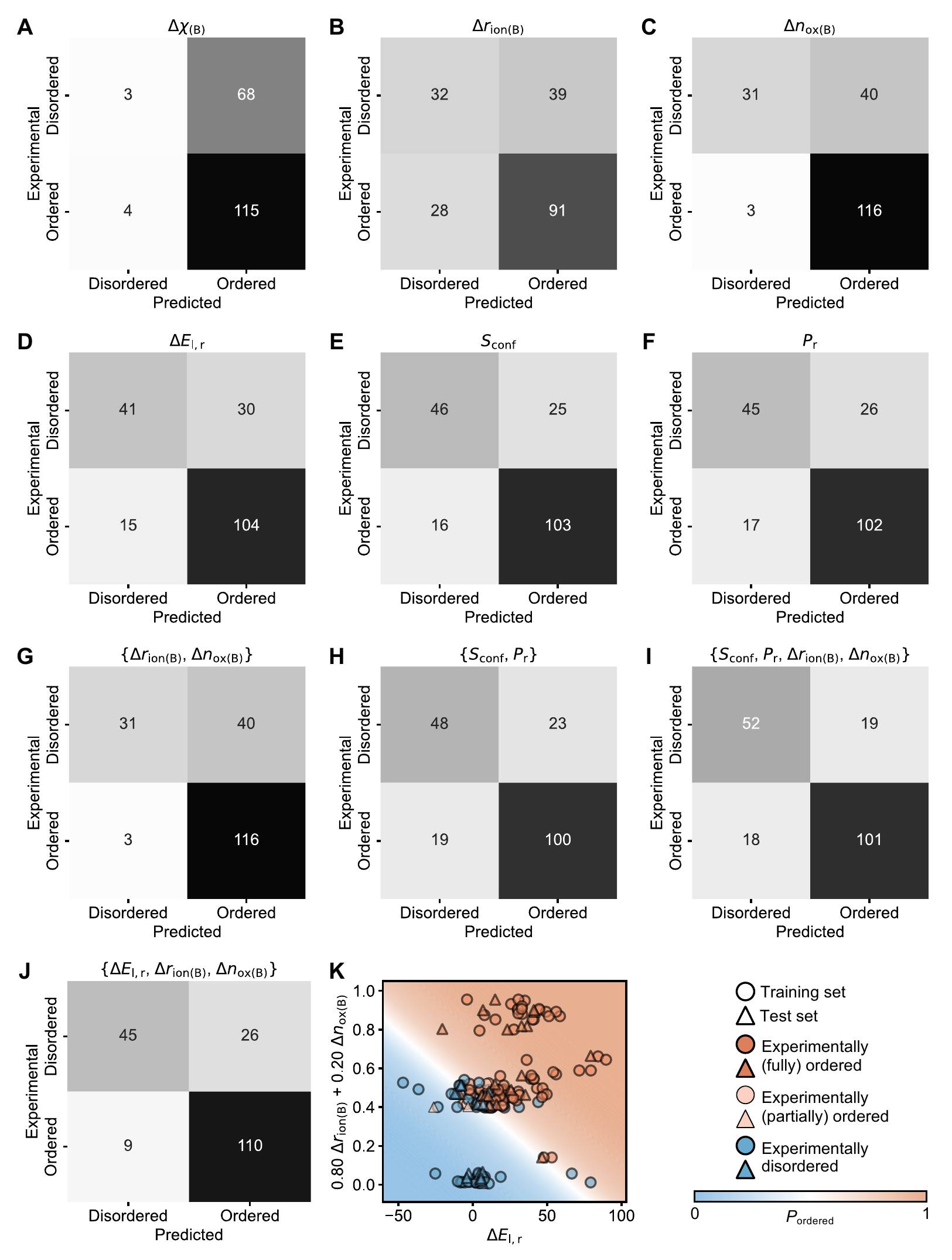}
\caption{
\textbf{Additional performance.}
(\textbf{A} to \textbf{J}) Confusion matrices for classifying $190$ A\textsubscript{2}B'B"O\textsubscript{6} perovskites as cation-ordered and disordered using logistic regression with five-fold cross-validation. $\Delta \chi_{\mathrm{(B)}}$, $\Delta r_{\mathrm{ion(B)}}$, and $\Delta n_{\mathrm{ox(B)}}$ are the differences between the electronegativities, ionic radii, and oxidation states of B-site ions, respectively. $\Delta E_{\mathrm{l,r}}$, $S_{\mathrm{conf}}$, and $P_{\mathrm{r}}$ are the energy gap between the layer and rocksalt arrangements, the configuration entropy, and the likelihood of generating the rocksalt configuration, respectively. (\textbf{K}) Decision boundary of the three-dimensional descriptor $\{\Delta E_{\mathrm{l,r}}, \Delta r_{\mathrm{ion(B)}}, \Delta n_{\mathrm{ox(B)}}\}$. $P_{\mathrm{ordered}}$ is the predicted probability that an oxide should be classified as cation-ordered through logistic regression.
}
\label{fig:SI_additional_descriptor_performance}
\end{center}
\end{figure}

\addtocounter{sfigure}{1}
\begin{figure}
\phantomsection
\begin{center}
\includegraphics[max size={\textwidth}{\textheight}]{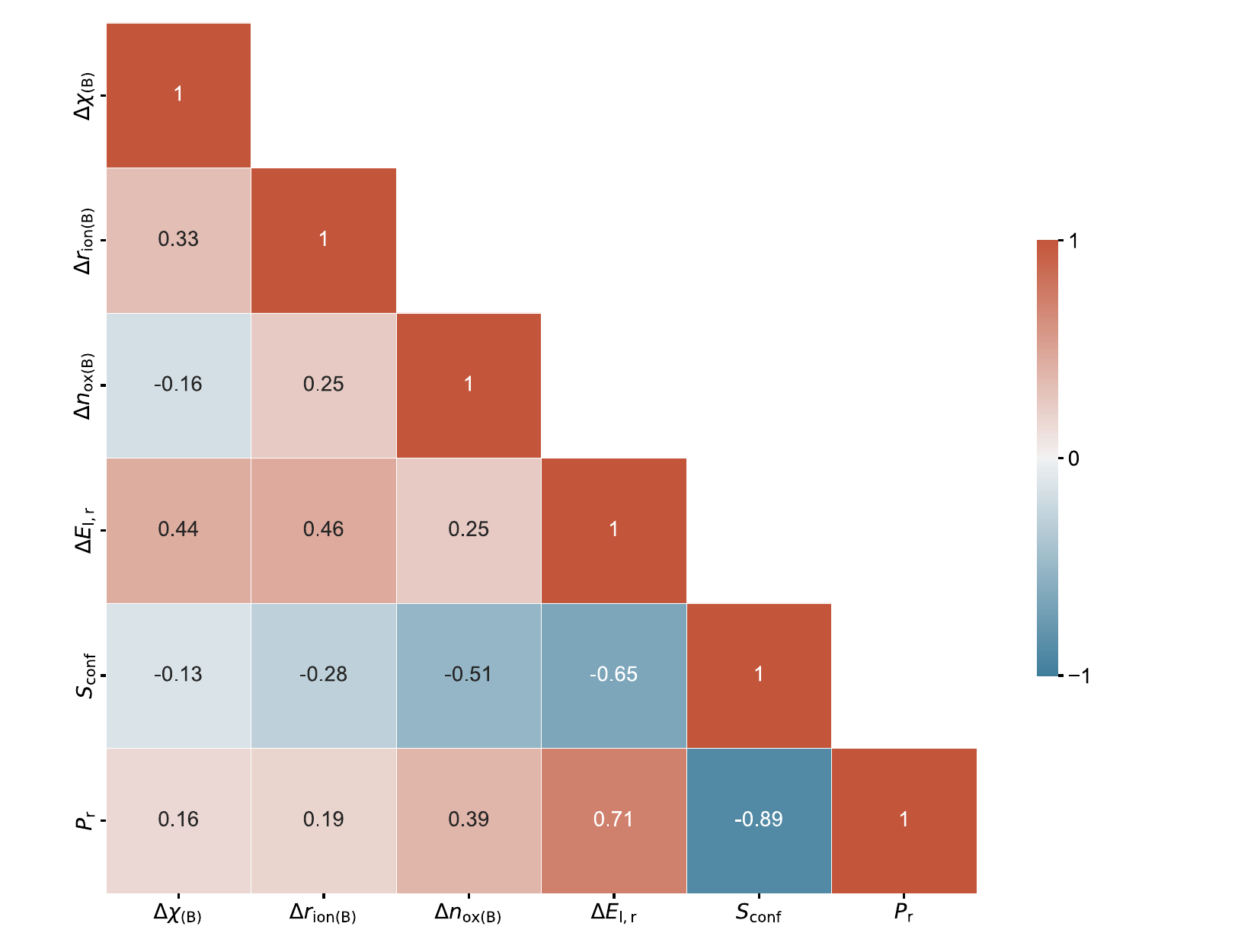}
\caption{
\textbf{Correlation matrix.}
Pearson correlation coefficients between different parameters for the experimental dataset of $190$ A\textsubscript{2}B'B"O\textsubscript{6} double perovskites. $\Delta \chi_{\mathrm{(B)}}$, $\Delta r_{\mathrm{ion(B)}}$, and $\Delta n_{\mathrm{ox(B)}}$ are the differences between the electronegativities, ionic radii, and oxidation states of B-site ions, respectively. $\Delta E_{\mathrm{l,r}}$, $S_{\mathrm{conf}}$, and $P_{\mathrm{r}}$ are the energy gap between the layer and rocksalt arrangements, the configuration entropy, and the likelihood of generating the rocksalt configuration, respectively.
}
\label{fig:SI_correlation_matrix}
\end{center}
\end{figure}

\addtocounter{sfigure}{1}
\begin{figure}
\phantomsection
\begin{center}
\includegraphics[max size={\textwidth}{\textheight}]{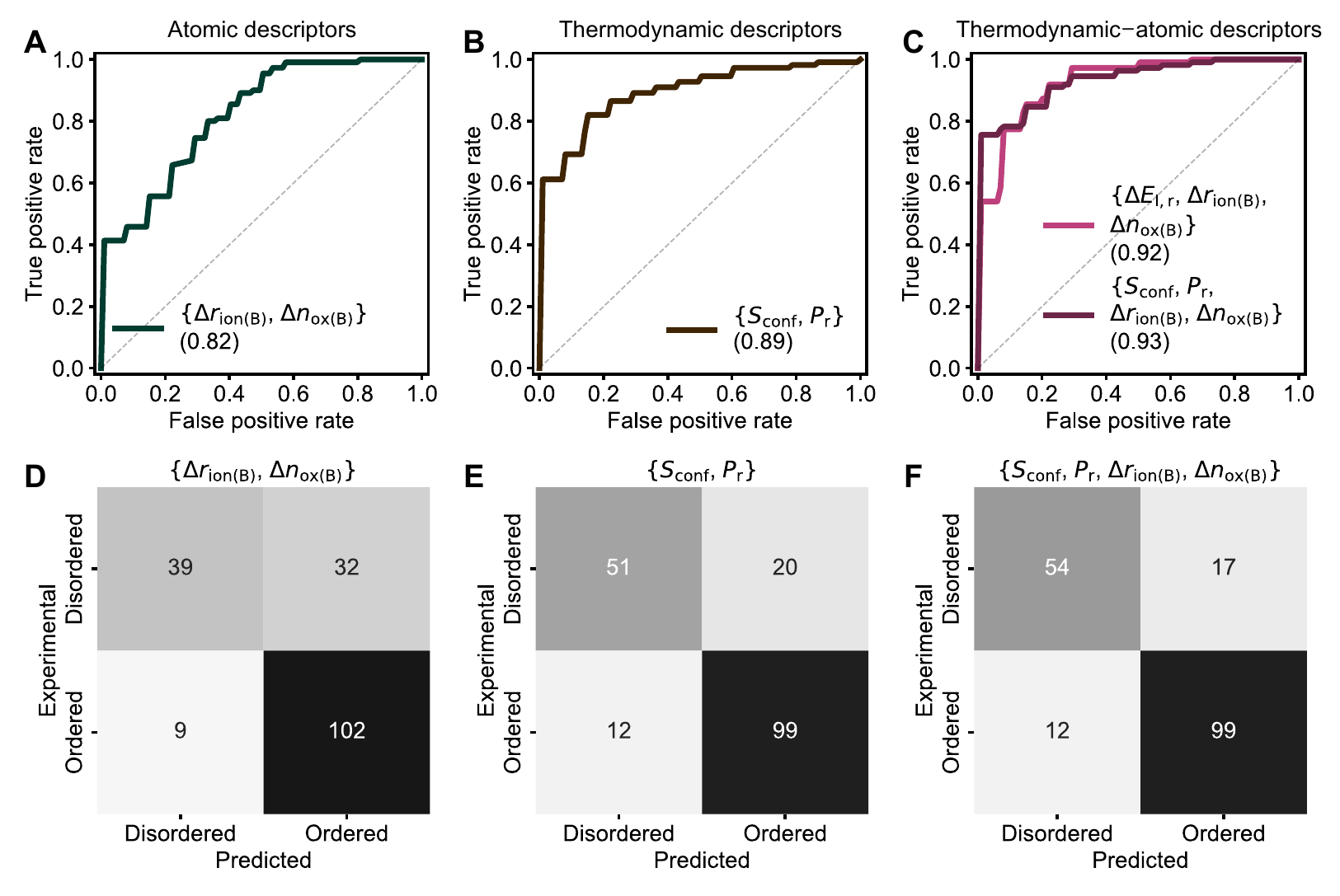}
\caption{
\textbf{Descriptor performance without experimentally partially cation-ordered oxides.}
(\textbf{A} to \textbf{C}) $182$ A\textsubscript{2}B'B"O\textsubscript{6} perovskites were left after filtering out eight experimentally partially cation-ordered oxides. Performance of multi-dimensional descriptors in classifying these $182$ oxides as cation-ordered and disordered using logistic regression with five-fold cross-validation, quantified by receiver operating characteristic (ROC) curves with area under the curve (AUC) in parentheses. Dashed lines show the ROC curves of a random classifier (AUC $= 0.5$). $\Delta \chi_{\mathrm{(B)}}$, $\Delta r_{\mathrm{ion(B)}}$, and $\Delta n_{\mathrm{ox(B)}}$ are the differences between the electronegativities, ionic radii, and oxidation states of B-site ions, respectively. $\Delta E_{\mathrm{l,r}}$, $S_{\mathrm{conf}}$, and $P_{\mathrm{r}}$ are the energy gap between the layer and rocksalt arrangements, the configuration entropy, and the likelihood of generating the rocksalt configuration, respectively. (\textbf{D} to \textbf{F}) Confusion matrices of top-performing multi-dimensional descriptors after filtering out eight experimentally partially cation-ordered oxides.
}
\label{fig:SI_filtering_partially_experimentally_ordered}
\end{center}
\end{figure}

\addtocounter{sfigure}{1}
\begin{figure}
\phantomsection
\begin{center}
\includegraphics[max size={\textwidth}{\textheight}]{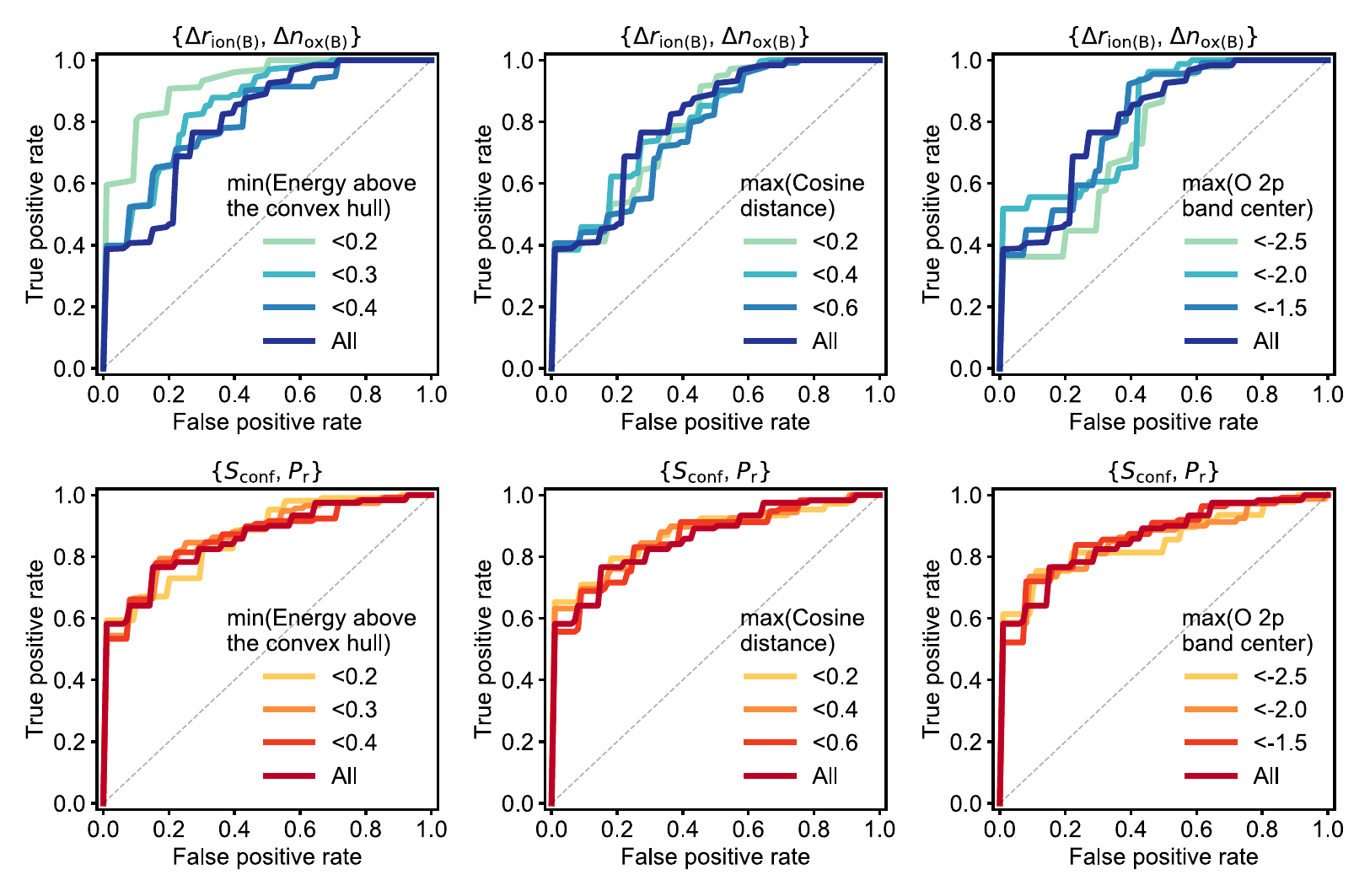}
\caption{
\textbf{Additional sensitivity analysis.}
ROC curves of top-performing two-dimensional descriptors in classifying $190$ A\textsubscript{2}B'B"O\textsubscript{6} perovskite oxides as cation-ordered or disordered before and after filtering them by limiting the range of selected density functional theory (DFT)--computed properties, including the minimum energy above the convex hull, the maximum cosine distance between the Matminer local order parameter fingerprint \cite{Ward:2018} of initialized and relaxed supercells, and the maximum O 2p band center relative to the Fermi level, which capture their tendency for phase metastability \cite{Sun:2016}, structural distortion \cite{Law:2023}, and oxygen nonstoichiometry \cite{Giordano:2022}, respectively. $\Delta r_{\mathrm{ion(B)}}$ and $\Delta n_{\mathrm{ox(B)}}$ are the differences between the ionic radii and oxidation states of B-site ions, respectively. $S_{\mathrm{conf}}$ and $P_{\mathrm{r}}$ are the configuration entropy and the likelihood of generating the rocksalt configuration, respectively.
}
\label{fig:SI_sensitivity_analysis}
\end{center}
\end{figure}

\addtocounter{sfigure}{1}
\begin{figure}
\phantomsection
\begin{center}
\includegraphics[max size={\textwidth}{\textheight}]{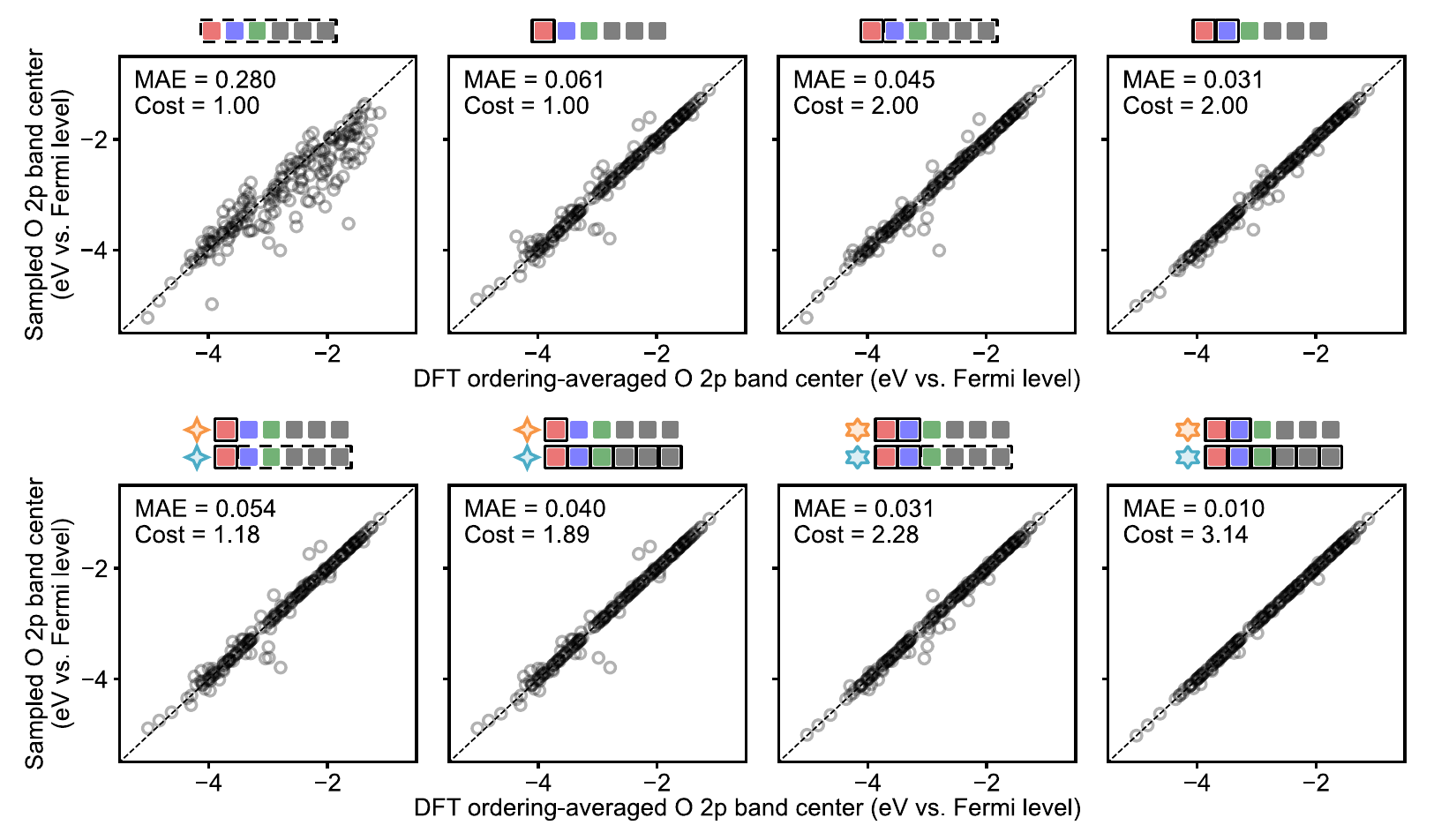}
\caption{
\textbf{Descriptor-driven and descriptor-free sampling of O 2p band center.}
Trade-offs between the accuracy and cost of different strategies for sampling the configurational space of cation arrangements in $190$ A\textsubscript{2}B'B"O\textsubscript{6} oxides. For each strategy, its computational cost was quantified based on the number of required DFT calculations per oxide, and its accuracy was obtained from the mean absolute error (MAE) of sampled properties with respect to the computationally expensive ground truth based on the ensemble average of all configurations. $\Delta r_{\mathrm{ion(B)}}$ and $\Delta n_{\mathrm{ox(B)}}$ are the differences between the ionic radii and oxidation states of B-site ions, respectively. $\Delta E_{\mathrm{l,r}}$ is the energy gap between the layer and rocksalt arrangements.
}
\label{fig:SI_selection_strategies_O2p}
\end{center}
\end{figure}

\addtocounter{sfigure}{1}
\begin{figure}
\phantomsection
\begin{center}
\includegraphics[max size={\textwidth}{\textheight}]{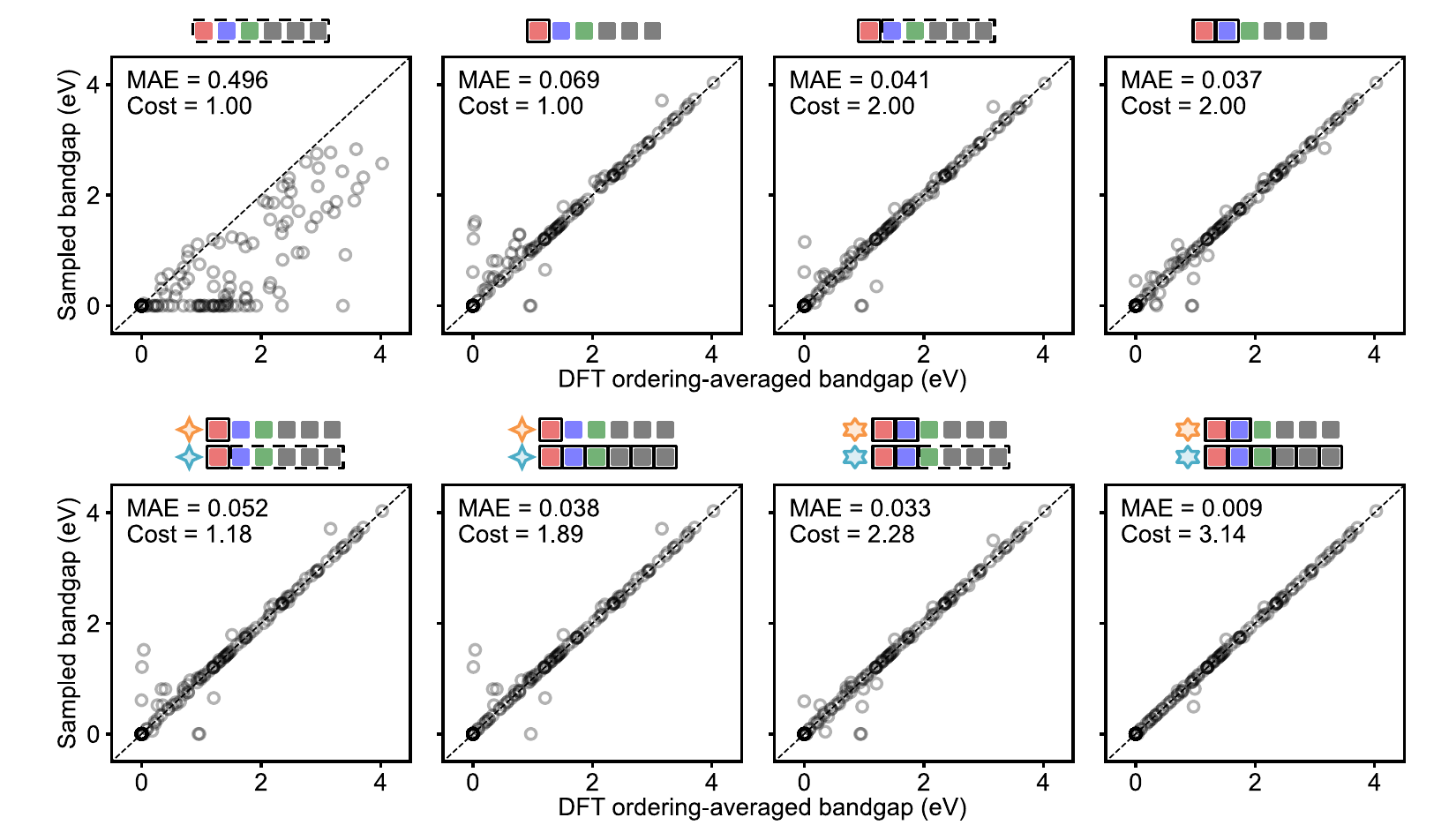}
\caption{
\textbf{Descriptor-driven and descriptor-free sampling of bandgap.}
Trade-offs between the accuracy and cost of different strategies for sampling the configurational space of cation arrangements in $190$ A\textsubscript{2}B'B"O\textsubscript{6} oxides. For each strategy, its computational cost was quantified based on the number of required DFT calculations per oxide, and its accuracy was obtained from the MAE of sampled properties with respect to the computationally expensive ground truth based on the ensemble average of all configurations. $\Delta r_{\mathrm{ion(B)}}$ and $\Delta n_{\mathrm{ox(B)}}$ are the differences between the ionic radii and oxidation states of B-site ions, respectively. $\Delta E_{\mathrm{l,r}}$ is the energy gap between the layer and rocksalt arrangements.
}
\label{fig:SI_selection_strategies_bandgap}
\end{center}
\end{figure}

\clearpage
\newcounter{SItable}
\renewcommand{\tablename}{Table}
\renewcommand{\thetable}{S\arabic{SItable}}

\addtocounter{SItable}{1}
\begin{table}
\phantomsection
\centering
\begin{threeparttable}
\caption{
\textbf{Atomic and thermodynamic primary features utilized for building descriptors using the sure independence screening and sparsifying operator (SISSO) method.}
}
\label{table:SI_sisso_definition}
\begin{tabular}{cl}
\hline
Feature & Definition \\
\hline
$\chi_{\mathrm{(A)}}$ & Pauling electronegativity of the A-site cations \\
$\Delta \chi_{\mathrm{(B)}}$ & Difference between the Pauling electronegativities of the B-site cations \\
$\bar{\chi}_{\mathrm{(B)}}$ & Average between the Pauling electronegativities of the B-site cations \\
$\chi_{\mathrm{(B')}} \chi_{\mathrm{(B'')}}$ & Product between the Pauling electronegativities of the B-site cations \\
$r_{\mathrm{ion(A)}}$ & Shannon effective ionic radius of the A-site cations\tnote{1} \\
$\Delta r_{\mathrm{ion(B)}}$ & Difference between the Shannon effective ionic radii of the B-site cations\tnote{1} \\
$\bar{r}_{\mathrm{ion(B)}}$ & Average between the Shannon effective ionic radii of the B-site cations\tnote{1} \\
$r_{\mathrm{ion(B')}} r_{\mathrm{ion(B'')}}$ & Product between the Shannon effective ionic radii of the B-site cations\tnote{1} \\
$n_{\mathrm{ox(A)}}$ & Nominal oxidation state of the A-site cations\\
$\Delta n_{\mathrm{ox(B)}}$ & Difference between the nominal oxidation states of the B-site cations \\
$\bar{n}_{\mathrm{ox(B)}}$ & Average between the nominal oxidation states of the B-site cations \\
$n_{\mathrm{ox(B')}} n_{\mathrm{ox(B'')}}$ & Product between the nominal oxidation states of the B-site cations \\
\hline
$\Delta E_{\mathrm{l,r}}$ & Energy gap between the layer and rocksalt arrangements\tnote{2} \\
$S_{\mathrm{conf}}$ & Configurational entropy of all possible cation arrangements\tnote{3} \\
$P_{\mathrm{r}}$ & Thermodynamic likelihood of forming the rocksalt arrangement \\
\hline
\end{tabular}
\begin{tablenotes}
\small
\item[1] Normalized by the ionic radius of an oxygen anion ($r_{\mathrm{ion(O^{2-})}} = 1.40\ \mathrm{\mathring{A}}$) to make it dimensionless.

\item[2] Normalized by $k_{\mathrm{B}}T$ to make it dimensionless, where $k_{\mathrm{B}}$ is the Boltzmann constant and $T$ is the temperature ($T = 1300\ \mathrm{K}$).

\item[3] Normalized by $\ln{N_{\mathrm{total}}}$ to make it dimensionless, where $N_{\mathrm{total}}$ is the total number of all possible cation configurations.
\end{tablenotes}
\end{threeparttable}
\end{table}

\addtocounter{SItable}{1}
\begin{table}
\phantomsection
\centering
\begin{threeparttable}
\caption{
\textbf{Importance of different features in constructing the best SISSO-learned one-dimensional atomic descriptor.}
}
\label{table:SI_sisso_importance_atomic}
\begin{tabular}{lcc}
\hline
Type & Feature & Relative importance\tnote{1} \\
\hline
\multirow{12}{11em}{All $12$ primary features} & $\Delta n_{\mathrm{ox(B)}}$ & $0.261$ \\
& $\bar{r}_{\mathrm{ion(B)}}$ & $0.114$ \\
& $\Delta \chi_{\mathrm{(B)}}$ & $0.105$ \\
& $r_{\mathrm{ion(B')}} r_{\mathrm{ion(B'')}}$ & $0.081$ \\
& $\bar{n}_{\mathrm{ox(B)}}$ & $0.081$ \\
& $n_{\mathrm{ox(A)}}$ & $0.064$ \\
& $\Delta r_{\mathrm{ion(B)}}$ & $0.047$ \\
& $n_{\mathrm{ox(B')}} n_{\mathrm{ox(B'')}}$ & $0.040$ \\
& $\bar{\chi}_{\mathrm{(B)}}$ & $0.037$ \\
& $\chi_{\mathrm{(B')}} \chi_{\mathrm{(B'')}}$ & $0.036$ \\
& $\chi_{\mathrm{(A)}}$ & $0.018$ \\
& $r_{\mathrm{ion(A)}}$ & $0.012$ \\
\hline
\multirow{12}{11em}{Top $12$ secondary features} & $\Delta n_{\mathrm{ox(B)}} + \Delta \chi_{\mathrm{(B)}}$ & $0.083$ \\
& $\Delta n_{\mathrm{ox(B)}} + \bar{r}_{\mathrm{ion(B)}}$ & $0.050$ \\
& $\Delta n_{\mathrm{ox(B)}} + \bar{n}_{\mathrm{ox(B)}}$ & $0.031$ \\
& $\bar{n}_{\mathrm{ox(B)}} + \Delta n_{\mathrm{ox(B)}}$ & $0.030$ \\
& $\Delta n_{\mathrm{ox(B)}} + r_{\mathrm{ion(B')}} r_{\mathrm{ion(B'')}}$ & $0.030$ \\
& $\Delta n_{\mathrm{ox(B)}} + \Delta r_{\mathrm{ion(B)}}$ & $0.025$ \\
& $\Delta n_{\mathrm{ox(B)}} - n_{\mathrm{ox(A)}}$ & $0.021$ \\
& $n_{\mathrm{ox(A)}} - \Delta n_{\mathrm{ox(B)}}$ & $0.020$ \\
& $\exp(\Delta n_{\mathrm{ox(B)}})$ & $0.015$ \\
& $\frac{\Delta n_{\mathrm{ox(B)}}}{n_{\mathrm{ox(A)}}}$ & $0.014$ \\
& $\Delta n_{\mathrm{ox(B)}} \bar{n}_{\mathrm{ox(B)}}$ & $0.013$ \\
& $\bar{n}_{\mathrm{ox(B)}} \Delta n_{\mathrm{ox(B)}}$ & $0.013$ \\
\hline
\end{tabular}
\begin{tablenotes}
\small
\item[1] Based on the number of occurrences in the $1000$ top-performing non-linear expressions minimizing the domain overlap between cation-disordered and ordered oxides identified by sure independence screening.  
\end{tablenotes}
\end{threeparttable}
\end{table}

\addtocounter{SItable}{1}
\begin{table}
\phantomsection
\centering
\begin{threeparttable}
\caption{
\textbf{Importance of different features in constructing the best SISSO-learned one-dimensional thermodynamic--atomic descriptor.}
}
\label{table:SI_sisso_importance_thermo_atomic}
\begin{tabular}{lcc}
\hline
Type & Feature & Relative importance\tnote{1} \\
\hline
\multirow{15}{11em}{All $15$ primary features} & $\Delta n_{\mathrm{ox(B)}}$ & $0.278$ \\
& $S_{\mathrm{conf}}$ & $0.142$ \\
& $P_{\mathrm{r}}$ & $0.092$ \\
& $\bar{r}_{\mathrm{ion(B)}}$ & $0.070$ \\
& $r_{\mathrm{ion(B')}} r_{\mathrm{ion(B'')}}$ & $0.052$ \\
& $\Delta \chi_{\mathrm{(B)}}$ & $0.038$ \\
& $\Delta r_{\mathrm{ion(B)}}$ & $0.037$ \\
& $\bar{n}_{\mathrm{ox(B)}}$ & $0.035$ \\
& $n_{\mathrm{ox(A)}}$ & $0.027$ \\
& $\bar{\chi}_{\mathrm{(B)}}$ & $0.017$ \\
& $\chi_{\mathrm{(B')}} \chi_{\mathrm{(B'')}}$ & $0.016$ \\
& $n_{\mathrm{ox(B')}} n_{\mathrm{ox(B'')}}$ & $0.015$ \\
& $\Delta E_{\mathrm{l,r}}$ & $0.014$ \\
& $\chi_{\mathrm{(A)}}$ & $0.009$ \\
& $r_{\mathrm{ion(A)}}$ & $0.007$ \\
\hline
\multirow{15}{11em}{Top $15$ secondary features} & $\Delta n_{\mathrm{ox(B)}} - S_{\mathrm{conf}}$ & $0.061$ \\
& $\Delta n_{\mathrm{ox(B)}} + P_{\mathrm{r}}$ & $0.059$ \\
& $\Delta n_{\mathrm{ox(B)}} + \Delta \chi_{\mathrm{(B)}}$ & $0.035$ \\
& $\Delta n_{\mathrm{ox(B)}} + \bar{r}_{\mathrm{ion(B)}}$ & $0.034$ \\
& $\exp(\Delta n_{\mathrm{ox(B)}})$ & $0.027$ \\
& $\Delta n_{\mathrm{ox(B)}} + r_{\mathrm{ion(B')}} r_{\mathrm{ion(B'')}}$ & $0.023$ \\
& $\Delta r_{\mathrm{ion(B)}} \Delta E_{\mathrm{l,r}}$ & $0.020$ \\
& $\Delta n_{\mathrm{ox(B)}}^4$ & $0.019$ \\
& $\Delta n_{\mathrm{ox(B)}}^3$ & $0.019$ \\
& $\Delta n_{\mathrm{ox(B)}}^2$ & $0.018$ \\
& $\Delta n_{\mathrm{ox(B)}} \Delta n_{\mathrm{ox(B)}}$ & $0.018$ \\
& $\Delta n_{\mathrm{ox(B)}} + \Delta r_{\mathrm{ion(B)}}$ & $0.017$ \\
& $\frac{\bar{r}_{\mathrm{ion(B)}}}{S_{\mathrm{conf}}}$ & $0.016$ \\
& $\frac{r_{\mathrm{ion(B')}} r_{\mathrm{ion(B'')}}}{S_{\mathrm{conf}}}$ & $0.016$ \\
& $\Delta n_{\mathrm{ox(B)}} + \Delta n_{\mathrm{ox(B)}}$ & $0.014$ \\
\hline
\end{tabular}
\begin{tablenotes}
\small
\item[1] Based on the number of occurrences in the $1000$ top-performing non-linear expressions minimizing the domain overlap between cation-disordered and ordered oxides identified by sure independence screening.  
\end{tablenotes}
\end{threeparttable}
\end{table}

\addtocounter{SItable}{1}
\begin{table}
\phantomsection
\centering
\begin{threeparttable}
\caption{
\textbf{Performance of the SISSO method with different settings.}
}
\label{table:SI_sisso_performance}
\begin{tabular}{lclc}
\hline
Type & Dimension & Method & ROC--AUC \\
\hline
\multirow{8}{10em}{Atomic} & \multirow{4}{1em}{$1$} & Convex hull, non-weighted & $0.8635$ \\
& & Convex hull, weighted & $0.8644$ \\
& & Decision tree, non-weighted & $0.8068$ \\
& & Decision tree, weighted & $0.8172$ \\
& \multirow{4}{1em}{$2$} & Convex hull, non-weighted & $0.8117$ \\
& & Convex hull, weighted & $0.8291$ \\
& & Decision tree, non-weighted & $0.8044$ \\
& & Decision tree, weighted & $0.8032$ \\
\hline
\multirow{8}{10.em}{Thermodynamic--atomic} & \multirow{4}{1em}{$1$} & Convex hull, non-weighted & $0.9211$ \\
& & Convex hull, weighted & $0.9432$ \\
& & Decision tree, non-weighted & $0.9424$ \\
& & Decision tree, weighted & $0.9435$ \\
& \multirow{4}{1em}{$2$} & Convex hull, non-weighted & $0.9115$ \\
& & Convex hull, weighted & $0.9307$ \\
& & Decision tree, non-weighted & $0.9197$ \\
& & Decision tree, weighted & $0.9442$ \\
\hline
\end{tabular}
\begin{tablenotes}
\small
\end{tablenotes}
\end{threeparttable}
\end{table}